\documentclass[journal,twoside,web]{ieeecolor}
\usepackage{tmi}
\usepackage{cite}
\usepackage{amsmath,amssymb,amsfonts}
\usepackage{algorithmic}
\usepackage{graphicx}
\usepackage[svgnames]{xcolor}

\usepackage[ruled,vlined]{algorithm2e}

\SetKwComment{Comment}{/* }{ */}
\newtheorem{theorem}{Theorem}

\usepackage{textcomp}
\usepackage{booktabs}
\usepackage{multirow}
\usepackage{amssymb}
\usepackage{caption}
\usepackage{subcaption}
\usepackage{hyperref}
\usepackage{soul}
\usepackage[normalem]{ulem}
\usepackage{caption}

\def\BibTeX{{\rm B\kern-.05em{\sc i\kern-.025em b}\kern-.08em
    T\kern-.1667em\lower.7ex\hbox{E}\kern-.125emX}}
\begin{document}
\title{CortexODE: Learning Cortical Surface Reconstruction by Neural ODEs}
\author{Qiang Ma, Liu Li, Emma C. Robinson, Bernhard Kainz, \IEEEmembership{Senior member, IEEE},\\ Daniel Rueckert,  \IEEEmembership{Fellow, IEEE}, Amir Alansary
\thanks{© 2022 IEEE. Personal use of this material is permitted. Permission from IEEE must be obtained for all other uses, in any current or future media, including reprinting/republishing this material for advertising or promotional purposes, creating new collective works, for resale or redistribution to servers or lists, or reuse of any copyrighted component of this work in other works.}
\thanks{This work was supported in part by the President’s PhD Scholarships at Imperial College London.}
\thanks{Q. Ma, L. Li, B. Kainz, D. Rueckert, and A. Alansary are with the BioMedIA group, Department of Computing, Imperial College London, UK (e-mail: q.ma20@imperial.ac.uk).}
\thanks{E. C. Robinson and B. Kainz are with King's College London, UK.}
\thanks{B. Kainz is also with Friedrich-Alexander-Universität Erlangen-Nürnberg (FAU), Germany.}
\thanks{D. Rueckert is also with Institute for AI and Informatics, in Medicine, Klinikum rechts der Isar, Technical University of Munich, Germany.}
}


\maketitle

\begin{abstract}
We present CortexODE, a deep learning framework for cortical surface reconstruction. CortexODE leverages neural ordinary differential equations (ODEs) to deform an input surface into a target shape by learning a diffeomorphic flow. The trajectories of the points on the surface are modeled as ODEs, where the derivatives of their coordinates are parameterized via a learnable Lipschitz-continuous deformation network. This provides theoretical guarantees for the prevention of self-intersections. CortexODE can be integrated to an automatic learning-based pipeline, which reconstructs cortical surfaces efficiently in less than 5 seconds. The pipeline utilizes a 3D U-Net to predict a white matter segmentation from brain Magnetic Resonance Imaging (MRI) scans, and further generates a signed distance function that represents an initial surface. Fast topology correction is introduced to guarantee homeomorphism to a sphere. Following the isosurface extraction step, two CortexODE models are trained to deform the initial surface to white matter and pial surfaces respectively. The proposed pipeline is evaluated on large-scale neuroimage datasets in various age groups including neonates (25-45 weeks), young adults (22-36 years) and elderly subjects (55-90 years). Our experiments demonstrate that the CortexODE-based pipeline can achieve less than 0.2mm average geometric error while being orders of magnitude faster compared to conventional processing pipelines.

\end{abstract}

\begin{IEEEkeywords}
Brain MRI, cortical surface reconstruction, geometric deep learning, neural ODE
\end{IEEEkeywords}

\section{Introduction}\label{sec:introduction}

\IEEEPARstart{T}{he} rapid advance of Magnetic Resonance Imaging (MRI)  has encouraged data acquisition for large cohort neuroimage studies, \emph{e.g.} ADNI~\cite{jack2008adni}, ABIDE~\cite{di2014abide}, OASIS~\cite{marcus2007oasis}, UK Biobank~\cite{sudlow2015ukbiobank}, the Human Connectome Project (HCP)~\cite{van2013wuminhcp}, and developing HCP (dHCP)~\cite{hughes2017dhcp,makropoulos2018dhcp}. To extract meaningful information from the resulting flood of data, various data processing pipelines have been developed for large-scale image analysis. One essential avenue for neuroimage studies is surface-based analysis of the cerebral cortex, which relies on accurate cortical surface reconstruction~\cite{dale1999cortical,fischl1999cortical,fischl2000measuring}.
Cortical surface reconstruction aims to visualize and analyze the 3D structure of the cerebral cortex by extracting its inner and outer surfaces from brain MRI scans~\cite{dale1999cortical,fischl1999cortical}. The inner cortical surface (or white matter surface) lies between the cortical gray matter (cGM) and white matter (WM) of the brain, and the outer surface (or pial surface) lies between the cerebrospinal fluid (CSF) and the cGM. As the cerebral cortex is highly folded, it is challenging to extract geometrically accurate cortical surface meshes. Naturally a cortical surface mesh must be a closed manifold and topologically equivalent to a 2-sphere, \emph{i.e.}, without any holes or handles.

\begin{figure*}[ht!]
\centering
\includegraphics[width=0.95\linewidth]{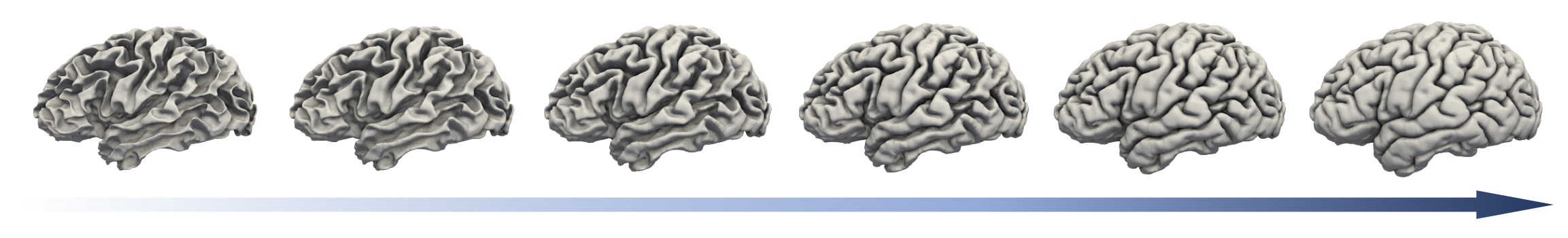}
\caption{A diffeomorphic flow modeled by CortexODE. A WM surface (leftmost) is deformed into a pial surface (rightmost).}
\label{fig:diffeo}
\end{figure*}

\subsection{Traditional Processing Pipelines}
Traditional cortical surface reconstruction approaches generally employ an empirically defined automatic processing pipeline. For example, the well-known neuroimage analysis software, FreeSurfer~\cite{fischl2012freesurfer}, reconstructs WM surfaces by applying mesh tessellation to the segmented WM, followed by mesh smoothing and topology correction~\cite{segonne2007topology}. The WM surfaces are then expanded to pial surfaces along with self-intersections checking. Although many neuroimage processing pipelines have been proposed and widely used, such as FreeSurfer~\cite{fischl2012freesurfer}, BrainSuite~\cite{shattuck2002brainsuite}, the HCP structural pipeline~\cite{glasser2013hcp} and the dHCP pipeline~\cite{makropoulos2018dhcp}, they are united by two major limitations.

Firstly, these pipelines are usually time-consuming as they integrate several computationally intensive geometric or image processing algorithms. As an example, FreeSurfer needs 4-8 hours to reconstruct the cortical surfaces for a single subject~\cite{fischl2012freesurfer}. This makes scaling to large neuroimage population studies challenging. 
Secondly, it is difficult to develop a single pipeline that can robustly process brain images from different age groups. For instance, neonatal brain images differ considerably in intensity values, size and shape from adults~\cite{allen2002volume,orasanu2014neonate,hughes2017dhcp,makropoulos2018dhcp}. This leads to poor reconstruction results when methods designed for adult neuroimage analysis are applied to such younger groups.

\subsection{Learning-based Cortical Surface Reconstruction}
As an end-to-end alternative to traditional pipelines, deep learning approaches have shown great potential for the reconstruction of 3D shapes while exhibiting convincingly fast runtime and powerful representation learning abilities. 

An increasing number of deep neural networks (DNNs) have been developed to tackle cortical surface reconstruction~\cite{henschel2020fastsurfer,cruz2020deepcsr,wickramasinghe2020voxel2mesh,ma2021pialnn,hong2021vox2surf,gopinath2021segrecon,lebrat2021corticalflow}. For instance, FastSurfer~\cite{henschel2020fastsurfer} reduced the processing time of the original FreeSurfer pipeline by adopting a convolutional neural network (CNN) for brain tissue segmentation and a fast spherical mapping for cortical surface reconstruction. Cruz et al.~\cite{cruz2020deepcsr} presented the DeepCSR framework to predict implicit surface representations involving occupancy fields and signed distance functions (SDFs). Both FastSurfer and DeepCSR rely on topology correction~\cite{bazin2005topology,bazin2007topology,segonne2007topology} to guarantee the genus-0 topology of the surfaces. They use the Marching Cubes algorithm~\cite{lorensen1987marching} for isosurface extractions. However, topology correction algorithms are computationally intensive, leading to $>$30 minutes of total runtime for both methods. In addition, the surfaces extracted from the volumetric representations are easily corrupted by partial volume effects~\cite{ballester2002partial}. 

Instead of learning an implicit surface, explicit learning-based approaches~\cite{wickramasinghe2020voxel2mesh,ma2021pialnn} aim to deform an initial mesh with pre-defined topology to a target mesh. To extract surfaces from 3D medical images, Wickramasinghe et al.~\cite{wickramasinghe2020voxel2mesh} proposed Voxel2Mesh to learn iterative mesh deformations. 
Several regularization terms were introduced to improve the mesh quality. Ma et al.~\cite{ma2021pialnn} proposed the PialNN architecture to expand an initial WM surface to a pial surface according to the local information captured from brain MRI volumes. These explicit approaches are fast since no topology correction is required. Nevertheless, the main issue is that there are no theoretical guarantees to prevent surface self-intersections.

\subsection{Learning Diffeomorphic Surface Deformation}
Non-manifold surfaces with self-intersections can lead to inaccurate estimation of morphological features of the cerebral cortex, such as cortical thickness, curvature, surface area and volume~\cite{fischl2000measuring}. Diffeomorphic transformation~\cite{ruelle1975diffeo,arsigny2004lie} is one effective way to preserve the manifoldness of the surfaces. In neuroimage analysis, the diffeomorphic transformation has been widely used for image registration and many frameworks have been presented such as LDDMM~\cite{beg2005lddmm} and DARTEL~\cite{ashburner2007dartel}. For learning-based approaches, Dalca et al. proposed VoxelMorph \cite{balakrishnan2019voxelmorph,dalca2019unsupervised} framework to learn stationary velocity field for medical image registration. The deformation field is solved by scaling and squaring method~\cite{arsigny2006squaring}.

To learn diffeomorphic surface deformation, Gupta and Chandraker~\cite{gupta2020nmf} proposed Neural Mesh Flow (NMF) with which they introduced neural ordinary differential equations (ODEs)~\cite{chen2018node} for 3D mesh generation. NMF learns a sequence of diffeomorphic flows between two meshes and models the trajectories of the mesh vertices as ODEs. This implicitly regularizes the manifold properties and reduces self-intersections in the generated meshes. However, NMF is limited by its insufficient shape representation capacity since it only augments a global feature vector for all vertices. 
Recently, Lebrat et al.~\cite{lebrat2021corticalflow} presented the CorticalFlow framework for cortical surface reconstruction. CorticalFlow employs diffeomorphic mesh deformation (DMD) modules to learn a series of stationary velocity fields, which define diffeomorphic deformations from an initial template to WM and pial surfaces. The forward Euler method is adopted to approximate the flow ODE and the numerical condition is derived to constrain the step size of the integration. This produces manifold meshes that avoid self-intersections effectively.

\subsection{Contributions}
In this work, we propose a novel geometric deep learning model for cortical surface reconstruction, called CortexODE, based on the neural ODE framework~\cite{chen2018node}. CortexODE learns continuous dynamics to deform an input surface to a target shape based on volumetric input. The trajectory of each vertex in the surface mesh is modeled as an ODE by a learnable deformation network, which constructs a diffeomorphic flow between the surfaces (see Figure~\ref{fig:diffeo}). The deformation network is proved to be Lipschitz continuous and the upper bound of the Lipschitz constant is estimated. Theoretically, this prevents surface self-intersections~\cite{ruelle1975diffeo,chen2018node,gupta2020nmf}. We further provide numerical conditions for general Runge-Kutta method to ensure each integration step is a homeomorphism.

We incorporate CortexODE into a learning-based processing pipeline. Given an input brain MRI volume, our pipeline employs a 3D U-Net~\cite{ronneberger2015unet} to predict a WM segmentation, which is transformed to an SDF by a distance transformation method~\cite{butt1998cdt}. To obtain an initial closed manifold surface, the SDF is processed by a topology correction algorithm~\cite{bazin2007topology}, which is re-implemented and improved to run within one second. An initial WM surface is extracted by the Marching Cubes algorithm~\cite{lorensen1987marching}. Then, two CortexODE models learn to shrink the initial surface to a WM surface and expand the WM surface to a pial surface respectively.
An entire CortexODE-based pipeline only needs 5 seconds to extract the cortical surfaces of both brain hemispheres. The pipeline is able to make continuous prediction in long-term neuroimage studies after appropriate training. We evaluate performance on publicly available large-scale neuroimage datasets in different age groups including dHCP, HCP and ADNI. The CortexODE code is open source and publicly available at \url{https://github.com/m-qiang/CortexODE}.

The main contributions of this work can be summarized as:
\begin{itemize}
\item The CortexODE-based pipeline can extract high quality cortical surfaces from brain MRI volumes within 5 seconds. It achieves the highest geometric accuracy compared to other state-of-the-art learning-based frameworks.
\item CortexODE learns a diffeomorphic flow to deform an input surface to a target shape, which provides theoretical guarantees for avoiding self-intersections without additional regularization terms.
\item We re-implement and accelerate the standard topology correction algorithm~\cite{bazin2007topology} to run $20\times$ faster.
\item The CortexODE pipeline is evaluated on multiple large-scale neuroimage datasets in various age groups. The reliability and robustness of the proposed pipeline are validated by test-retest experiments and ablation studies.
\end{itemize}

\begin{figure*}[ht!]
\centering
\includegraphics[width=0.9\linewidth]{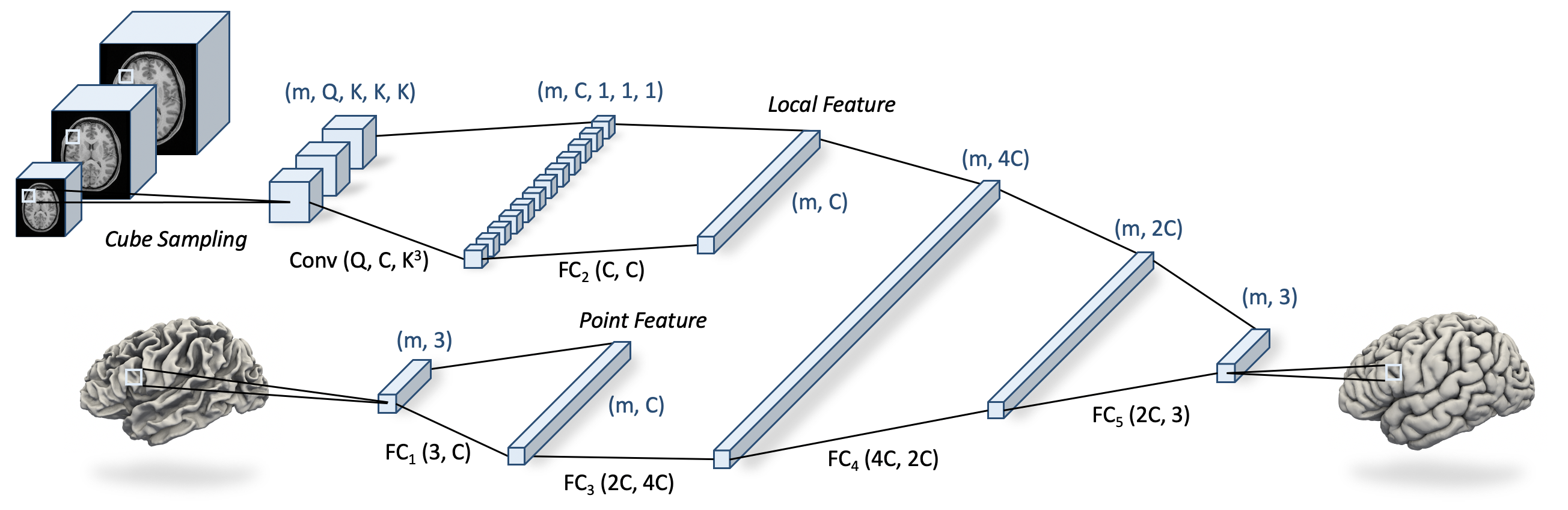}
\caption{The architecture of the deformation network in the CortexODE framework. The deformation network extracts a point feature from the coordinate of the surface point, and extracts a local feature from multi-scale brain MRI inputs by cube sampling. Two feature vectors are concatenated to predict the derivatives of the points. Here $m$ is the number of the input surface points, $C$ is the dimension of the feature vectors, and $Q$ is the number of the input scales.}
\label{fig:deform-net}
\end{figure*}

\section{CortexODE Framework}\label{sec:cortexode}
In this section, we introduce the CortexODE framework. 
Given an input surface, CortexODE models the trajectories of the points on the surface as ODEs, where the derivatives of the points are parameterized by a DNN called \emph{deformation network}. We show that the deformation network is Lipschitz continuous with respect to the input points. This provides theoretical guarantees for the prevention of self-intersections according to the existence and uniqueness theorem for the solutions of ODE initial value problems (IVPs)~\cite{coddington1955ode}.

\subsection{Neural ODE-based Framework}
\subsubsection{Neural ODEs}
A neural ODE \cite{chen2018node} can be defined as:
\begin{equation}\label{eq:neural-ode}
\frac{\mathrm{d}\mathbf{x}(t)}{\mathrm{d}t} = F_{\theta}(\mathbf{x}(t),t),
\end{equation}
where the right-hand side (RHS) $F_{\theta}$ is a DNN with parameters $\theta$. $\mathbf{x}(t)\in\Omega$ is a continuous hidden state of the neural network for $t\in[0, T]$, and $\Omega$ is an arbitrary domain such as color images or $\mathbb{R}^n$ vectors. The neural ODE takes $\mathbf{x}(0)=\mathbf{x}_0$ as the input while the output is the final state $\mathbf{x}(T)$. The unique solution of the ODE~(\ref{eq:neural-ode}) exists if the RHS $F_{\theta}$ is Lipschitz continuous~\cite{coddington1955ode}, \emph{i.e.}, there exists a constant $L_F>0$ such that for any $\mathbf{x},\mathbf{y}\in\Omega$,
\begin{equation}
\|F_{\theta}(\mathbf{x})-F_{\theta}(\mathbf{y})\|_p \leq L_{F}\|\mathbf{x}-\mathbf{y}\|_p,
\end{equation}
where $\|\mathbf{x}\|_p=(\sum_i|x_i|^p)^{\frac{1}{p}}$ is the $p$-norm for $p\geq1$.

\subsubsection{CortexODE}
Given an initial surface $\mathcal{S}_0\subset\mathbb{R}^3$, based on neural ODEs, CortexODE models the trajectory of each point on the surface as an autonomous ODE:
\begin{equation}\label{eq:cortexode}
\frac{\mathrm{d}\mathbf{x}(t)}{\mathrm{d}t} = F_{\theta}(\mathbf{x}(t),\mathbf{V}),~\mathbf{x}(0)=\mathbf{x}_0,
\end{equation}
where $\mathbf{x}_0\in\mathcal{S}_0$ is the coordinate of a point on the surface $\mathcal{S}_0$. The solution $\mathbf{x}(t)=\mathbf{x}_0+\int_0^t F_{\theta}(\mathbf{x}(s),\mathbf{V})\mathrm{d}s$ represents the trajectory of each point for $t\in[0,T]$. $\mathbf{x}(T)$ is the final coordinate after deformation. $\mathbf{V}\in\mathbb{R}^{D_1\times D_2\times D_3}$ is a volumetric input, and $F_{\theta}:\mathbb{R}^3\times\mathbb{R}^{D_1\times D_2\times D_3}\rightarrow\mathbb{R}^3$ is a learnable \emph{deformation network} parameterizing the derivatives of the coordinates as shown in Figure~\ref{fig:deform-net}.

CortexODE learns a continuously deformable surface $\mathcal{S}(t)$ with points $\mathbf{x}(t)\in\mathcal{S}(t)$ for $t\in[0,T]$. A diffeomorphic flow~\cite{ruelle1975diffeo,gupta2020nmf} is defined by the ODE in Eq.~(\ref{eq:cortexode}) from the initial surface $\mathcal{S}(0)=\mathcal{S}_0$ to the final prediction $\mathcal{S}(T)$. According to the existence and uniqueness theorem for ODE solutions~\cite{coddington1955ode}, if $F_{\theta}$ is Lipschitz continuous with respect to $\mathbf{x}$, the trajectories $\mathbf{x}(t)$ of different points will never intersect with each other, which prevents the surface from intersecting with itself.

\subsection{Deformation Network}\label{sec:deform-net}
The deformation network $F_{\theta}$, \emph{i.e.}, the RHS of Eq.~(\ref{eq:cortexode}), follows the architecture of PialNN~\cite{ma2021pialnn} and is designed to be Lipschitz continuous. The architecture is shown in Figure~\ref{fig:deform-net}.

\subsubsection{Point Feature}
Given $m$ points on the input surface $\mathcal{S}$, we use a fully connected (FC) neural network layer $f_1$ to extract the \emph{point feature} $\mathbf{z}_{point}=f_1(\mathbf{x})\in\mathbb{R}^C$ from the coordinate $\mathbf{x}\in\mathbb{R}^3$ of each point, where $C$ is the dimension of the feature vector $\mathbf{z}_{point}$. The $l$th FC layer is defined as $f_l(\mathbf{x})=\phi(W_l\mathbf{x}+\mathbf{b}_l)$, where $\phi$ is a non-linear activation function, $W_l$ and $\mathbf{b}_l$ are learnable weight and bias.

\subsubsection{Cube Sampling}
Next, the local information of each surface point is captured from the brain MRI intensity. As shown in Figure~\ref{fig:deform-net}, for each surface point $\mathbf{x}$, we find its location in the multi-scale brain MRI inputs. For each scale, a cube of size $K^3$ is extracted, which contains the voxel at the location $\mathbf{x}$ as well as its neighborhood voxels in the brain MRI. This procedure is called \emph{cube sampling}.

More specifically, for a brain MRI volume $\mathbf{V}$ with size of $D_1\times D_2\times D_3$, we consider multi-scale inputs $\mathbf{V}_q$ with size of $\lfloor\frac{D_1}{2^{q-1}}\rfloor\times \lfloor\frac{D_2}{2^{q-1}}\rfloor \times \lfloor\frac{D_3}{2^{q-1}}\rfloor$, which downsamples the original image by $1/2^{q-1}$ for $q\in\{1,..,Q\}$, where $Q$ is the number of the input scales. Then for each coordinate $\mathbf{x}$, we can find its corresponding location in all volumes $\mathbf{V}_q$. The voxel value $v_q\in\mathbb{R}$ at the location $\mathbf{x}$ in the $q$th-scale brain MRI input $\mathbf{V}_q$ is computed by a trilinear interpolation $v_q=\mathrm{Lerp}(\mathbf{x}/2^{q-1},\mathbf{V}_q)$, where $\mathrm{Lerp}(\cdot)$ denotes the interpolation function. The coordinate $\mathbf{x}$ is scaled by $1/2^{q-1}$ to match the location in $\mathbf{V}_q$. To capture the neighborhood information of $\mathbf{x}$, for each scale $q$, we consider a grid of size $K^3$ with center $\mathbf{x}/2^{q-1}$, which is defined by a set $\{\mathbf{x}/2^{q-1}+\delta_{ijk}\}_{i,j,k\in\{1,..,K\}}$ for $\delta_{ijk}:=\left[i-\lceil K/2\rceil,j-\lceil K/2\rceil,k-\lceil K/2\rceil\right]\in\mathbb{Z}^3$. Therefore, for each scale $q$, we can sample a cube $\mathbf{v}_q$ of size $K^3$, containing the voxel values at the locations of all points in the grid. Letting $\mathbf{v}(\mathbf{x})\in\mathbb{R}^{Q\times K\times K\times K}$ be the stack of all cubes $\mathbf{v}_q$ at point $\mathbf{x}$, then each entry of $\mathbf{v}$ can be defined as:
\begin{equation}\label{eq:v_qijk}
\mathbf{v}_{qijk}(\mathbf{x})=\mathrm{Lerp}(\mathbf{x}/2^{q-1}+\delta_{ijk},\mathbf{V}_q),
\end{equation}
for $1\leq q\leq Q$ and $i,j,k\in\{1,...,K\}$.

\subsubsection{Local Feature}
After cube sampling, we obtain a total of $m\cdot Q$ cubes. As shown in Figure \ref{fig:deform-net}, a convolutional layer with kernel size $K$ followed by a FC layer $f_2$ is employed to encode the cubes $\mathbf{v}(\mathbf{x})$ to a \emph{local feature} $\mathbf{z}_{local}\in\mathbb{R}^C$ for each point $\mathbf{x}\in\mathcal{S}$. Here the convolutional layer is equivalent to a FC layer $f_0$, so that we have $\mathbf{z}_{local}=f_2\circ f_0(\mathbf{v}(\mathbf{x}))$.

\subsubsection{Deformation}
We concatenate the point feature and local feature of $\mathbf{x}$ to a new feature vector, which is passed to a sequence of FC layers $f_3,f_4,f_5$ to learn the final deformation $F_{\theta}(\mathbf{x},\mathbf{V})\in\mathbb{R}^3$. Finally, the deformation network $F_{\theta}$ can be expressed explicitly as:
\begin{equation}\label{eq:F}
F_{\theta}(\mathbf{x},\mathbf{V})=f_5\circ f_4\circ f_3\left([f_1(\mathbf{x})^{\top},f_2\circ f_0(\mathbf{v}(\mathbf{x}))^{\top}]^{\top}\right).
\end{equation}

\subsection{Existence and Uniqueness of Solutions}
In this section, we show the deformation network $F_\theta$ is Lipschitz continuous, so that there exits a unique solution for the IVP (\ref{eq:cortexode}). Assuming the activation functions $\phi$ in $F_\theta$ are ReLU or LeakyReLU, we have the following theorem.
\begin{theorem}\label{thm1}
The RHS function $F_{\theta}$ of the ODE in Eq.~(\ref{eq:cortexode}) is Lipschitz continuous with respect to the input $\mathbf{x}\in\mathbb{R}^3$ with Lipschitz constant
\begin{equation}\label{eq:L}
\begin{split}
L_F=\|W_5\|_p&\|W_4\|_p\|W_3\|_p\big(\|W_1\|_p^p\\
&+2K^3(\mathbf{V}_{\mathrm{max}}-\mathbf{V}_{\mathrm{min}})^p\|W_2\|_p^p\|W_0\|_p^p\big)^{\frac{1}{p}}.
\end{split}
\end{equation}
Then, there exists a unique solution for the IVP in Eq.~(\ref{eq:cortexode}).
\end{theorem}
\begin{proof}
See Appendix \ref{proof:thm1}.
\end{proof}
Here $\mathbf{V}_{\max}$ and $\mathbf{V}_{\min}$ denote the maximum and minimum entry of $\mathbf{V}$. Based on Theorem~\ref{thm1}, for different points $\mathbf{x}_0\in\mathcal{S}_0$, the solutions $\mathbf{x}(t)$ are unique and their trajectories will never intersect with each other. This theoretically prevents self-intersections in surface $\mathcal{S}(t)$ for $t\in[0,T]$.

\begin{figure*}[ht]
\centering
\includegraphics[width=0.98\linewidth]{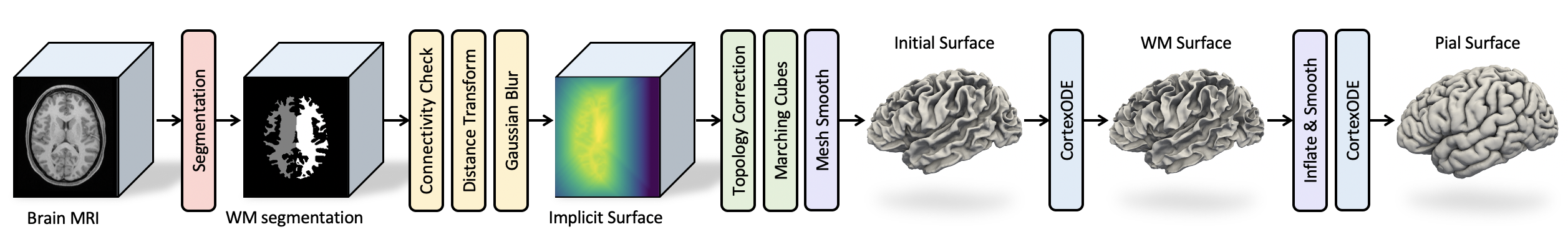}
\caption{The CortexODE-based cortical surface reconstruction pipeline. Given a brain MRI volume, the pipeline learns to predict a WM segmentation and then transforms it to a signed distance function (SDF). After topology correction and isosurface extraction, two CortexODE models are trained to deform an initial surface into a WM surface and to deform an inflated WM surface into a pial surface respectively.}
\label{fig:pipeline}
\end{figure*}

\subsection{CortexODE Discretization}\label{sec:ode-integrate}

To solve the IVP for Eq.~(\ref{eq:cortexode}) in CortexODE, we use explicit fixed-step methods for ODE discretization:
\begin{equation}
\mathbf{x}_{n+1} = \mathbf{x}_{n} + h\Delta\mathbf{x}_{n},
\end{equation}
where $n\in\{0,...,N-1\}$, $N$ is the number of the total steps of the iteration, and $h=T/N$ is the step size. To estimate $\Delta\mathbf{x}_{n}$, \cite{lebrat2021corticalflow} adopted explicit forward Euler method. In this work, we consider an $r$-stage explicit Runge-Kutta (RK) method~\cite{butcher1964rk}. The general form of RK method for an autonomous ODE can be formulated as:
\begin{equation}\label{eq:rk}
\mathbf{x}_{n+1} = \mathbf{x}_{n} + h\sum\nolimits_{i=1}^r\beta_i s_i(\mathbf{x}_n),
\end{equation}
where $s_i$ is defined by
\begin{equation}
s_i(\mathbf{x})=F_{\theta}\Big(\mathbf{x}+h\sum\nolimits_{j=1}^{i-1}\alpha_{i,j}s_j(\mathbf{x})\Big).
\end{equation} 
The coefficients $\beta_i$ and $\alpha_{i,j}$ are usually defined by a Butcher tableau~\cite{butcher1964rk}. Next we show that under certain conditions, each integration step in (\ref{eq:rk}) is a homeomorphism, which can reduce surface self-intersections numerically.

\begin{theorem}\label{thm2}
Let $\mathcal{G}(\mathbf{x})=\mathbf{x}+h\sum_{i=1}^r\beta_i s_i(\mathbf{x})$ where $r$ is the stage of the RK method. If the time step $h$ and the Lipschitz constant $L$ of $F_{\theta}$ satisfy $\eta(h,L)<1$, where 
\begin{equation}\label{eq:eta}
\eta(h,L):=\sum_{c=1}^{r}h^cL^c\sum_{i_0=1}^r|\beta_{i_0}|\prod_{k=1}^{c-1}\sum_{i_k=1}^{i_{k-1}-1}|\alpha_{i_{k-1},i_k}|,
\end{equation}
then $\mathcal{G}$ is a homeomorphism, \emph{i.e.}, $\mathcal{G}$ is a continuous bijection and $\mathcal{G}^{-1}$ is continuous.
\end{theorem}
\begin{proof}
See Appendix \ref{proof:thm2}.
\end{proof}

In this work, we consider forward Euler, Midpoint, and 4th-order RK methods as ODE solvers. Theorem \ref{thm2} extends the results of homeomorphism from an Euler method \cite{lebrat2021corticalflow} to an any-stage RK method. The Euler method can be viewed as a special case of the RK method with $\beta_1=1$ and $\alpha_{i,j}=0$. Then the sufficient condition of homeomorphism is $\eta(h,L)=hL<1$. Similarly, for 2nd-order Midpoint method with $\beta_1=0$, $\beta_2=1$ and $\alpha_{2,1}=1/2$, it should hold $\eta(h,L)=hL+\frac{1}{2}h^2L^2<1$. In practice, to reduce the training time, we use a larger step size $h$ (\emph{e.g.} $h=0.1$) to solve the ODEs during training. In the inference phase, we can use a smaller $h$ to reduce surface self-intersections introduced by the numerical errors. (See Section~\ref{sec:solver} for detailed comparison and discussion.)

By solving the ODE, CortexODE deforms an initial surface $S(0)= S_0$ to the final predicted surface $S(T)=S_N$ with points $\mathbf{x}_N$. According to Theorem \ref{thm2}, the surface self-intersections can be prevented numerically as each integration step is a bijective mapping. More specifically, for each integration step, any two distinct points on the surface cannot be mapped to the same position.
However, it should be noted that for a polygon mesh representation, we only deform the vertices of the mesh instead of all points on the surface. As a result, the limited mesh resolution, \emph{i.e.}, the number of vertices, can also introduce a few mesh self-intersections. (See Section~\ref{sec:solver}.)

\section{CortexODE-based Processing Pipeline}\label{sec:pipeline}
In this section, we introduce a CortexODE-based automatic processing pipeline for cortical surface reconstruction, which is depicted in Figure~\ref{fig:pipeline}. 

\subsection{White Matter Segmentation}
Given an input brain MRI volume $\mathbf{V}\in\mathbb{R}^{D_1\times D_2\times D_3}$, the CortexODE-based pipeline uses a deep neural network to learn a WM segmentation map $\mathbf{U}\in[0,1,2]^{D_1\times D_2\times D_3}$. The segmentation has three categories involving background, left WM and right WM. Note that here the WM segmentation refers to the interior of the WM surface, which contains cortical WM, deep gray matter, ventricle, hippocampus and other tissues within the surface. To predict the segmentation, a 3D U-Net~\cite{ronneberger2015unet} is trained by minimizing the cross entropy loss $\mathcal{L}_{ce}$ between the prediction and ground truth (GT) segmentations. The GT segmentations are generated by traditional pipelines~\cite{hughes2017dhcp,glasser2013hcp,fischl2012freesurfer} which leverage the anatomical information of the brain MRI. More advanced segmentation approaches, such as QuickNAT~\cite{roy2019quicknat} and FastSurfer~\cite{henschel2020fastsurfer}, can be used to improve the performance of the segmentation and the following surface reconstruction (see Section \ref{sec:ablation-seg}).

\subsection{Signed Distance Transformation}
As shown in Figure~\ref{fig:pipeline}, based on the predicted WM segmentation $\mathbf{U}$, we generate a signed distance function (SDF) which represents an implicit surface.

\begin{figure}[ht]
\centering
\includegraphics[width=0.8\linewidth]{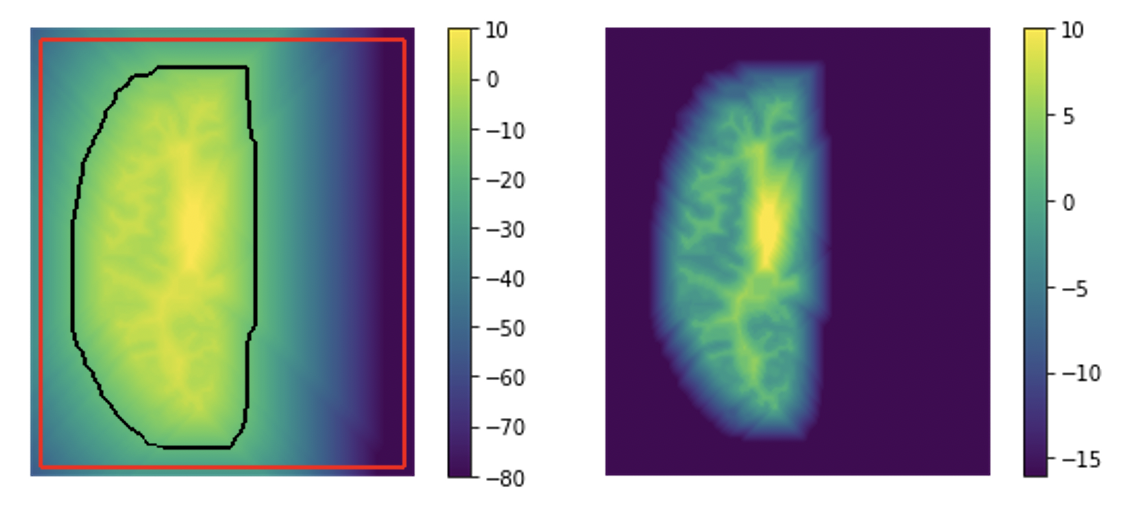}
\vspace{10pt}
\caption{Left: A discrete SDF. The red bounding box marks the initial points for the original topology correction algorithm~\cite{bazin2007topology}. The black contour marks the initial points on the boundary $\mathcal{S}^{\hat{\lambda}}$ for our improved version. Right: The topology correction result with boundary $\mathcal{S}^{\hat{\lambda}=-16}$.}
\label{fig:sdf}
\end{figure}

We first check the connectivity of the binary WM segmentation and select the largest connected component which contains most of the voxels. Next, a distance transform algorithm~\cite{butt1998cdt} is adopted to transform the segmentation $\mathbf{U}$ to a volumetric level set $\mathcal{U}_d\in\mathbb{R}^{D_1\times D_2\times D_3}$, where the value of each voxel in $\mathcal{U}_d$ is the distance from its location to the boundary of the WM segmentation as shown in Figure~\ref{fig:sdf}. The voxels in the exterior of the segmentation have negative value while the interior voxel values are positive. A Gaussian blur with standard deviation $\sigma$ is  used to smooth the level set.

Then, a continuous SDF $\mathcal{U}:\mathbb{R}^3\rightarrow\mathbb{R}$ can be obtained by applying a trilinear interpolation $\mathrm{Lerp}(\cdot)$ to the level set $\mathcal{U}_d$, that is, $\mathcal{U}(\cdot)=\mathrm{Lerp}(\cdot,\mathcal{U}_d)$. Each level $\lambda\in\mathbb{R}$ in the SDF defines an implicit surface:
\begin{equation}\label{eq:sdf}
\mathcal{S}^{\lambda}=\{\mathbf{x}~|~\mathcal{U}(\mathbf{x})=\lambda\},
\end{equation}
where $\mathbf{x}\in\mathbb{R}^3$ are the points on the surfaces. The 0-level surface $\mathcal{S}^0$ defines an implicit WM surface.

\subsection{Topology Correction}
To guarantee spherical topology of the implicit surfaces, we apply the topology correction algorithm by Bazin et al.~\cite{bazin2007topology} to the discrete SDF $\mathcal{U}_d$. The algorithm~\cite{bazin2007topology} adopts a topology-preserving fast marching approach to fix the topology of the surfaces $\mathcal{S}^{\lambda}$ defined by $\mathcal{U}$ under any level $\lambda$. However, since the original implementation~\cite{bazin2007topology,cruz2020deepcsr} is computationally intensive, we re-implement and improve the original algorithm, which is consequently accelerated by a factor of $20$.

First of all, we re-implement the topology correction algorithm~\cite{bazin2007topology} in Python, while the original algorithm was implemented by Java. We use Numba~\cite{lam2015numba} to accelerate the algorithm. Numba is a just-in-time (JIT) compiler for Python, which is suitable for dense computation.

In addition, we improve the initialization of the original algorithm~\cite{bazin2007topology}. To fill the holes or handles in the input level set, the original algorithm~\cite{bazin2007topology} starts topology correction from initial points on the boundary of the level set, which is shown as the red bounding box in Figure~\ref{fig:sdf}. The algorithm processes the voxels iteratively from lower level to higher level. Therefore, only the voxels inside the boundary need to be processed. For cortical surface extraction, since we consider one brain hemisphere at a time, instead of correcting the topology of the entire level set, we only need to process the voxels inside a $\hat{\lambda}$-level surface $\mathcal{S}^{\hat{\lambda}}$ for $\hat{\lambda}<0$. This not only accelerates the runtime but also ensures that the 0-level surface, \emph{i.e.}, the WM surface, has desired topology.

To obtain the boundary points of the surface $\mathcal{S}^{\hat{\lambda}}$, we first create a binary mask containing all voxels inside the surface $\mathcal{S}^{\hat{\lambda}}$. More precisely, the voxel of the binary mask is set to one if its level $\lambda$ satisfies $\lambda\geq\hat{\lambda}$.
Then, we apply morphological dilation to enlarge and expand the binary mask. The dilated voxels represent the boundary of $\mathcal{S}^{\hat{\lambda}}$, which is illustrated as the black contour in Figure~\ref{fig:sdf}. These boundary voxels are used as initial points by topology correction algorithm, and the values of the voxels outside the boundary are set to $\hat{\lambda}$. Figure~\ref{fig:sdf}-Right shows a topologically corrected result with $\hat{\lambda}=-16$. The initialization avoids processing of $76\%$ of the voxels. Such an accelerated topology correction algorithm only needs one second to process a volume of size $192\times224\times192$.

\subsection{Cortical Surface Reconstruction}
\subsubsection{Initial Surface Extraction}
We use the Marching Cubes algorithm~\cite{lorensen1987marching} to extract an initial WM surface $\mathcal{S}_0=\mathcal{S}^{\lambda_0}$ from the topology corrected SDF at the level $\hat{\lambda}<\lambda_0<0$, which is between the WM and pial surfaces. 
The initial surface $\mathcal{S}_0$ is represented by a 3D triangular mesh $\mathcal{M}_0$ with vertices $\mathbf{x}_0^i\in\mathcal{M}_0$ for $i\in\{1,...,m\}$, where $m$ is the number of vertices. The surface mesh $\mathcal{M}_0$ is smoothed twice by Laplacian smoothing $\bar{\mathbf{x}}_0^i=\sum\nolimits_{j\in\mathcal{N}(i)}{\mathbf{x}}_0^j/|\mathcal{N}(i)|$, where $\mathcal{N}(i)$ is the adjacency list of the $i$-th vertex defined by mesh $\mathcal{M}_0$.

\subsubsection{WM Surface Reconstruction}
As shown in Figure~\ref{fig:pipeline}, two CortexODE models learn to shrink the initial WM surface to a WM surface and to expand the WM surface to a pial surface respectively. To train the CortexODE models, we minimize the distance between predicted surface mesh $\hat{\mathcal{M}}$ and ground truth (GT) mesh $\mathcal{M}_*$ without regularization terms. We generate the GT surfaces by traditional pipelines~\cite{hughes2017dhcp,glasser2013hcp,fischl2012freesurfer}.

For WM surface reconstruction, the input for CortexODE is the initial WM surface generated by the pipeline. The bidirectional Chamfer distance~\cite{wang2020pixel2mesh, wickramasinghe2020voxel2mesh} is computed as the loss function $\mathcal{L}_{ch}$ between $\hat{\mathcal{M}}$ and $\mathcal{M}_*$:
\begin{equation}\label{eq:chamfer}
\mathcal{L}_{ch}=\sum_{\hat{\mathbf{x}}\in\hat{\mathcal{M}}}\min_{\mathbf{x}_*\in\mathcal{M}_*}\|\hat{\mathbf{x}}-\mathbf{x}_*\|^2 + 
\sum_{\mathbf{x}_*\in\mathcal{M}_*}\min_{\hat{\mathbf{x}}\in\hat{\mathcal{M}}}\|\hat{\mathbf{x}}-\mathbf{x}_*\|^2,
\end{equation}
where $\hat{\mathbf{x}}$ and $\mathbf{x}_*$ are the coordinates of the points sampled uniformly on the predicted and GT meshes. We sample 150k points on both meshes to compute the loss.

\subsubsection{Pial Surface Reconstruction}
Since it is challenging to reconstruct an accurate pial surface, we use an inflated WM surface as the input for CortexODE to extract pial surfaces. The inflating and smoothing process is given in Algorithm~\ref{algo:inflate}.
\begin{algorithm}
\caption{Mesh Inflating and Smoothing}
\label{algo:inflate}
\SetKwInput{kwData}{Input}
\kwData{A mesh $\mathcal{M}$ with $m$ vertices $\mathbf{x}^i$, the number of iterations $N$, and the scale parameter $\rho$.}
\For{$n\leq N$}{
\For{$i\leq m$}{
$\mathbf{x}^i=\sum_{j\in\mathcal{N}(i)}\mathbf{x}^j / |\mathcal{N}(i)|$ \Comment*[r]{Smooth} }
\For{$i\leq m$}{
Compute normals $\mathbf{n}^i$\;
$\mathbf{x}^i=\mathbf{x}^i+\rho\mathbf{n}^i$  \Comment*[r]{Inflate}
}}
\end{algorithm}

During training, we use inflated GT WM surfaces as the input for CortexODE, which have the same vertex connectivity as GT pial surfaces. Thus the predicted (or output) pial surfaces also have point-to-point correspondence to the GT pial surfaces in the training. Then, we can compute the mean squared error (MSE) loss $\mathcal{L}_{mse}$ between the predicted pial surfaces $\hat{\mathcal{M}}$ and GT pial surfaces $\mathcal{M}_*$ directly without point matching:
\begin{equation}\label{eq:mse}
\mathcal{L}_{mse}(\hat{\mathcal{M}},\mathcal{M}_*)=\frac{1}{m}\sum\nolimits_{i=1}^{m}\|\hat{\mathbf{x}}^i-\mathbf{x}_*^i\|^2,
\end{equation}
where $m$ is the number of mesh vertices.

\begin{table*}[t]
\centering
\caption{Comparative results of cortical surface reconstruction on geometric accuracy and self-intersections. The average symmetric surface distance (ASSD), Hausdorff distance (HD), and the ratio of the self-intersecting faces (SIF) are measured for WM and pial surfaces on both brain hemispheres. The mean value (M), standard deviation (SD), and $p$-value are reported. For all metrics lower mean values indicate better results. *Significant difference ($p<0.05$) compared to CortexODE.}
\label{table:comparison}\scriptsize
\setlength\tabcolsep{3.2pt} 
\begin{tabular}{cl|ccc|ccc|ccc|ccc|ccc|ccc}
\toprule
& \multicolumn{1}{c}{} &\multicolumn{9}{c}{WM Surface} &\multicolumn{9}{c}{Pial Surface} \\
\cmidrule(l{3.5pt}r{3.5pt}){3-11}\cmidrule(l{3.5pt}r{3.5pt}){12-20}
&\multicolumn{1}{l}{} & \multicolumn{3}{c}{ASSD ($mm$)} & \multicolumn{3}{c}{HD ($mm$)} & \multicolumn{3}{c}{SIF ($\%$)} & \multicolumn{3}{c}{ASSD ($mm$)} & \multicolumn{3}{c}{HD ($mm$)} & \multicolumn{3}{c}{SIF ($\%$)}\\
\midrule
& Method & M & SD & $p$ &  M & SD & $p$ &  M & SD & $p$ &
M & SD & $p$ & M & SD & $p$ & M & SD & $p$ \\
\midrule
\multirow{5}{*}{\rotatebox[origin=c]{90}{dHCP}}
& CortexODE & 
0.107 & 0.024 & --- & 0.210 & 0.073 & --- &
0.020 & 0.014 & --- & 0.128 & 0.027 & --- &
0.275 & 0.074 & --- & \textbf{0.284} & 0.147 & --- \\
& PialNN & 
\textbf{0.106} & 0.023 & 0.612 & \textbf{0.207} & 0.072 & 0.500 &
0.041 & 0.041 & $<$0.001* & \textbf{0.117} & 0.026 & $<$0.001* &
\textbf{0.245} & 0.069 & $<$0.001* & 9.401 & 1.992 & $<$0.001* \\
& CorticalFlow & 
0.142 & 0.035 & $<$0.001* & 0.285 & 0.133 & $<$0.001* &
0.073 & 0.056 & $<$0.001* & 0.172 & 0.041 & $<$0.001* &
0.436 & 0.174 & $<$0.001* & 8.627 & 2.367 & $<$0.001* \\
& DeepCSR & 
0.216 & 0.083 & $<$0.001* & 0.509 & 0.328 & $<$0.001* &
--- & --- & --- & 0.401 & 0.102 & $<$0.001* &
1.522 & 0.504 & $<$0.001* & --- & --- & --- \\
& Voxel2Mesh & 
0.262 & 0.040 &$<$0.001* & 0.557 & 0.102 & $<$0.001* &
\textbf{0.010} & 0.018 &$<$0.001* & 0.270 & 0.048 & $<$0.001* &
0.664 & 0.136 &$<$0.001* & 6.315 & 1.617 & $<$0.001* \\
\midrule
\multirow{5}{*}{\rotatebox[origin=c]{90}{HCP}}
& CortexODE & 
\textbf{0.115} & 0.011 & --- & \textbf{0.233} & 0.022 & --- &
\textbf{0.004} & 0.004 & --- & \textbf{0.163} & 0.018 & --- &
\textbf{0.347} & 0.040 & --- & \textbf{0.367} & 0.155 & --- \\
& PialNN & 
0.117 & 0.010 & 0.077 & 0.237 & 0.021 & 0.015* &
0.033 & 0.037 & $<$0.001* & 0.194 & 0.019 & $<$0.001* &
0.413 & 0.041 & $<$0.001* & 2.599 & 0.929 & $<$0.001* \\
& CorticalFlow & 
0.163 & 0.015 & $<$0.001* & 0.353 & 0.040 & $<$0.001* &
0.389 & 0.250 & $<$0.001* & 0.236 & 0.025 & $<$0.001* &
0.557 & 0.061 & $<$0.001* & 1.049 & 0.472 & $<$0.001* \\
& DeepCSR & 
0.201 & 0.034 & $<$0.001* & 0.424 & 0.079 & $<$0.001* &
--- & --- & --- & 0.310 & 0.052 & $<$0.001* &
0.827 & 0.223 & $<$0.001* & --- & --- & --- \\
& Voxel2Mesh & 
0.275 & 0.021 & $<$0.001* & 0.641 & 0.068 & $<$0.001* &
0.011 & 0.017 & $<$0.001* & 0.306 & 0.025 & $<$0.001* &
0.753 & 0.068 & $<$0.001* & 1.962 & 0.606 & $<$0.001* \\
\midrule
\multirow{5}{*}{\rotatebox[origin=c]{90}{ADNI}}
& CortexODE & 
0.197 & 0.046 & --- & 0.423 & 0.103 & --- &
\textbf{0.010} & 0.011 & --- & \textbf{0.186} & 0.049 & --- &
\textbf{0.399} & 0.094 & --- & 0.142 & 0.090 & --- \\
& PialNN & 
\textbf{0.194} & 0.046 & 0.468 & \textbf{0.417} & 0.104 & 0.463 &
0.086 & 0.084 & $<$0.001* & 0.206 & 0.042 & $<$0.001* &
0.438 & 0.087 & $<$0.001* & 2.975 & 1.441 & $<$0.001* \\
& CorticalFlow & 
0.254 & 0.050 & $<$0.001* & 0.587 & 0.139 & $<$0.001* &
0.055 & 0.055 & $<$0.001* & 0.238 & 0.044 & $<$0.001* &
0.557 & 0.124 & $<$0.001* & \textbf{0.064} & 0.037 & $<$0.001* \\
& DeepCSR & 
0.296 & 0.051 & $<$0.001* & 0.633 & 0.101 & $<$0.001* &
--- & --- & --- & 0.292 & 0.064 & $<$0.001* &
0.616 & 0.165 & $<$0.001* & --- & --- & --- \\
& Voxel2Mesh & 
0.372 & 0.048 & $<$0.001* & 0.907 & 0.151 & $<$0.001* &
0.033 & 0.036 & $<$0.001* & 0.344 & 0.043 & $<$0.001* &
0.802 & 0.115 & $<$0.001* & 0.493 & 0.314 & $<$0.001* \\
\bottomrule
\end{tabular}
\end{table*}

\section{Experiments}\label{sec:experiments}

\subsection{Experiment Details}
\subsubsection{Dataset} We evaluate the performance of our CortexODE pipeline on three publicly available datasets: the Alzheimer's Disease Neuroimaging Initiative (ADNI) dataset~\cite{jack2008adni}, the WU-Minn Human Connectome Project (HCP) Young Adult dataset~\cite{van2013wuminhcp}, and the third released developing HCP (dHCP) dataset~\cite{makropoulos2018dhcp}. We split each dataset into training, validation, and testing data by the ratio of 6/3/1.

Specifically, for the ADNI dataset, we use 524 T1-weighted (T1w) brain MRI from subjects aged from 55 to 90 years. The MRI volumes are aligned rigidly to the MNI152 template and clipped to the size of $176\times208\times176$ at $1mm^3$ isotropic resolution. The ground truth (GT) segmentations and surfaces are generated by FreeSurfer v7.2~\cite{fischl2012freesurfer}. For the HCP dataset, we use 600 T1w brain MRI from 22-36 year-old young adults. The size of the data is $192\times224\times192$ and the resolution is $1mm^3$. The GT data are produced by the HCP structural pipeline~\cite{glasser2013hcp}. The dHCP dataset is collected from neonates at age 25 to 45 weeks. We use 874 T2w MRI images, which are resampled to size of $160\times208\times208$ at $0.7mm^3$ isotropic resolution. The GT surfaces are extracted by the dHCP processing pipeline~\cite{makropoulos2018dhcp}. For all datasets, the GT WM surface mesh and pial surface mesh have the same vertex connectivity. The coordinates of the vertices are normalized to $[-1,1]$.

\subsubsection{Implementation Details} For initial surface generation, we train a U-Net for 200 epochs to predict the WM segmentations. After distance transform, the SDF is smoothed by a Gaussian filter with standard deviation $\sigma=0.5$. For topology correction, we set the level $\hat{\lambda}=-16$ for initialization. The initial WM surfaces are extracted at level $\lambda_0=-0.8$. The WM surfaces are inflated and smoothed twice with scale parameter $\rho=0.002$ (Algorithm~\ref{algo:inflate}) as the input to reconstruct pial surfaces.  

For CortexODE model, we use $Q=3$ scales for the input and the side length of the cubes is set to $K=5$. We set $C=128$ as the dimension of feature vectors. Our model is lightweight and it only contains 328,835 learnable parameters. The model is discretized by an Euler solver with total numerical steps $N=10$ and step size $h=0.1$ for training. For inference, we set $N=20$ for all datasets to reduce self-intersections.

We adopt the \emph{adjoint sensitivity method} proposed by Chen et al.~\cite{chen2018node} to train CortexODE models. This allows constant GPU memory costs during training despite the step size of ODE integration. We train four CortexODE models for 500 epochs to reconstruct WM and pial surfaces of both left and right brain hemispheres. The Adam optimizer is used with learning rate $10^{-4}$ and the batch size is set to 1. The training time and GPU memory costs are reported in Section \ref{sec:solver} for different ODE solvers. The entire CortexODE-based pipeline requires 5.29GB GPU memory for inference.

\subsubsection{Baselines} We compare our CortexODE-based pipeline with state-of-the-art (SOTA) deep learning approaches for cortical surface reconstruction, including PialNN~\cite{ma2021pialnn}, CorticalFlow~\cite{lebrat2021corticalflow}, DeepCSR~\cite{cruz2020deepcsr}, and Voxel2Mesh~\cite{wickramasinghe2020voxel2mesh}. Since PialNN is designed for pial surface reconstruction, we use initial WM surfaces generated by the CortexODE-based pipeline as the input of PialNN to reconstruct WM surfaces. The same initial WM surface meshes are simplified to 5k vertices as the input for Voxel2Mesh, as it is difficult to deform spheres to cortical surfaces. The input mesh is subdivided iteratively to 164k vertices by Voxel2Mesh. For CorticalFlow, we use subject-specific convex hulls as the input surfaces instead of a global convex hull which leads to unstable training in our experiments.
For DeepCSR, the predicted SDF is $2\times$ the size of the input MRI volume to reduce partial volume effects. For acceleration, we apply the same topology correction algorithm as used in our pipeline for DeepCSR. 
For CorticalFlow and DeepCSR, the reconstructed surface meshes have up to 500k vertices. For CortexODE and PialNN, the predicted meshes have approximately 120k--160k vertices for the HCP and ADNI datasets, as well as 60k--150k vertices for the dHCP dataset.
For fair comparison, all models including CortexODE are trained on an Nvidia RTX 3080 GPU with 12GB memory. 

\subsection{Cortical Surface Reconstruction Results} 
\begin{figure*}[ht]
\centering
\includegraphics[width=0.95\linewidth]{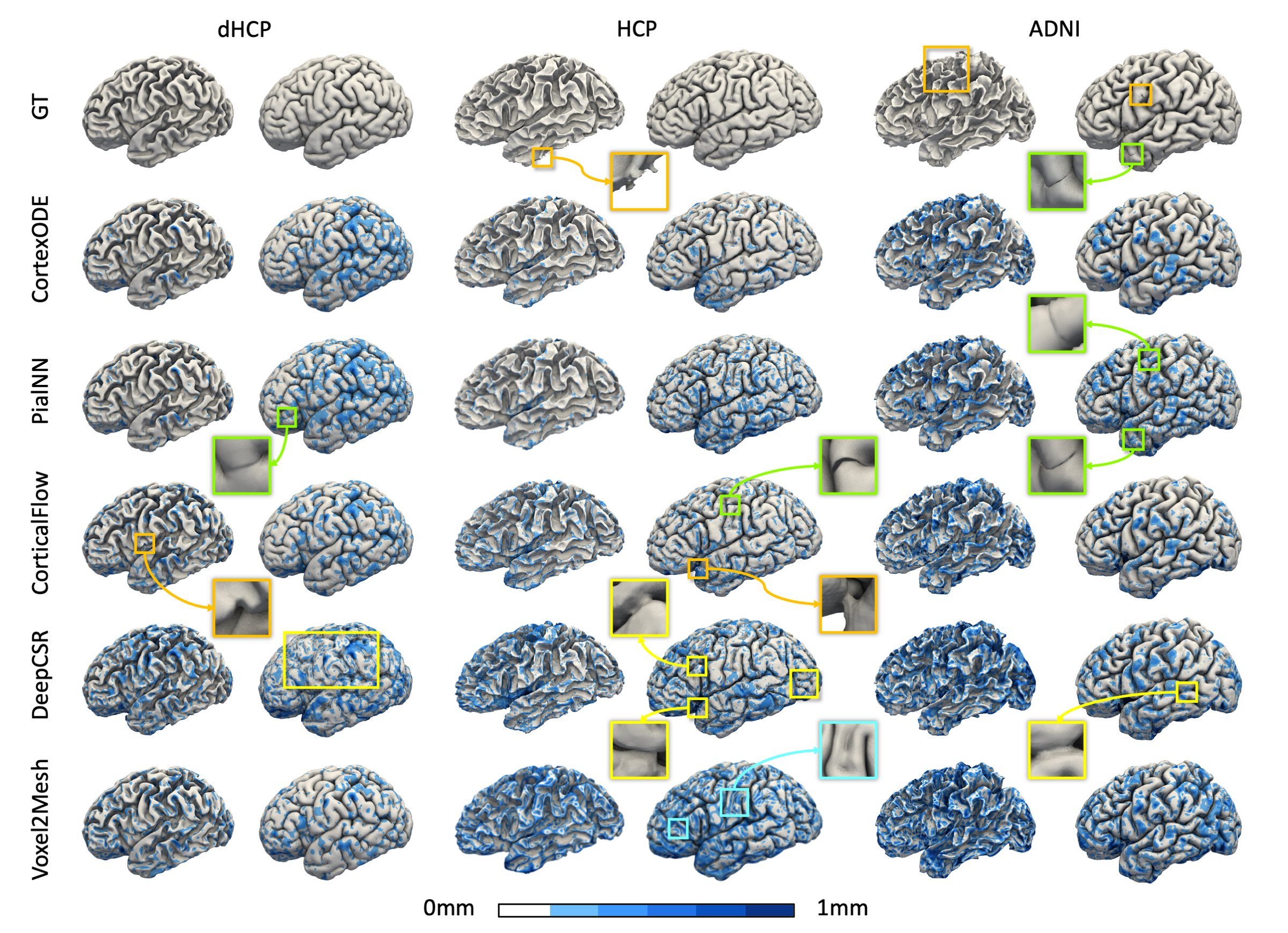}
\caption{Visualization of the reconstructed cortical surfaces for dHCP, HCP and ADNI datasets. The geometric error map (blue color) visualizes the distance (from $0$ to $1mm$) between the predicted surfaces and the ground truth. The artifacts ({\color{Orange}corruption}, {\color{LimeGreen}self-intersection}, {\color{Gold}partial volume effect}, and {\color{DeepSkyBlue}over-smoothness}) on the surfaces are highlighted by different colors.}
\label{fig:surface}
\end{figure*}

\begin{figure*}[ht]
\centering
\includegraphics[width=0.9\linewidth]{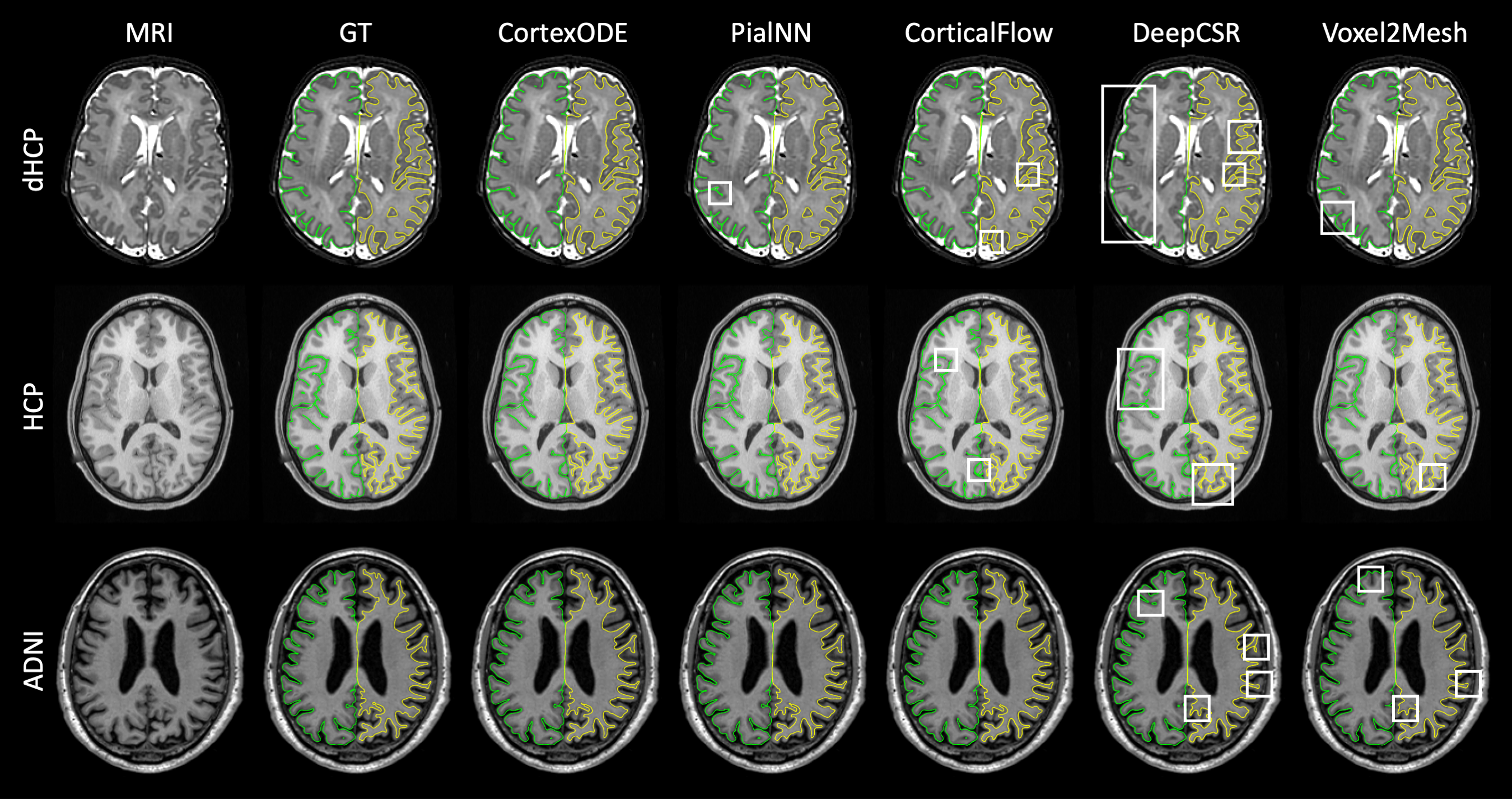}
\caption{Visualization of left pial surfaces (green) and right WM surfaces (yellow) projected on the MRI slices. The artifacts are highlighted by white boxes.}
\label{fig:mri}
\end{figure*}

We compare the geometric accuracy, self-intersections and runtime of CortexODE with SOTA deep learning baselines. The comparative results are given in Table~\ref{table:comparison} and Figure \ref{fig:surface}--\ref{fig:runtime}.
\subsubsection{Geometric Accuracy} We consider two distance-based metrics~\cite{cruz2020deepcsr} to measure geometric accuracy: average symmetric surface distance (ASSD) and Hausdorff distance (HD). The ASSD measures the average distance between two surfaces. It samples a batch of points on one surface and computes the mean distance between the points and the other surface, and vice versa. Similarly, the HD computes the maximum distance between two surfaces. For the purpose of robustness, the 90th percentile distance is computed rather than the maximum value~\cite{cruz2020deepcsr,lebrat2021corticalflow}.  We use PyTorch3D~\cite{ravi2020pytorch3d} to uniformly sample 100k points on the surface meshes to compute bidirectional ASSD and HD between the predicted and GT surfaces for all baselines. The mean and standard deviation of the geometric errors are reported in Table~\ref{table:comparison}. The lower distance means better geometric accuracy. 

Furthermore, we use an independent t-test to examine the statistical significance between CortexODE and other baseline models. The $p$-value of the t-test is reported in Table~\ref{table:comparison}. As common practice, we assume that $p<0.05$ indicates there is significant difference between the geometric errors of CortexODE and compared baseline. Table~\ref{table:comparison} shows that CortexODE and PialNN achieve better performance on both WM and pial surface reconstruction than other approaches. For WM surface reconstruction, since PialNN uses the initial surfaces generated by CortexODE pipeline, there is no significant difference between CortexODE and PialNN. For pial surfaces, CortexODE is significantly better than PialNN on the HCP and ADNI datasets. Although PialNN outperforms CortexODE on the dHCP dataset, it produces 9.4\% self-intersections which can lead to highly inaccurate estimation of cortical features (see Section \ref{sec:clinical-mesh}).

\subsubsection{Visual Inspection} We visualize the reconstructed cortical surfaces of the subjects in the dHCP, HCP and ADNI datasets. Figure~\ref{fig:surface} shows the cortical surface meshes with color maps reflecting their geometric errors, and Figure~\ref{fig:mri} depicts the projected surfaces on the brain MRI slices. The artifacts of the surfaces are highlighted. According to the visualization, the CortexODE-based pipeline produces surfaces with good mesh quality and high geometric accuracy. The results match the ground truth surfaces generated by the traditional pipelines, or even have better mesh quality. The pial surfaces reconstructed by PialNN is not smoothed well and more self-intersections are observed as highlighted. The DeepCSR results are corrupted by partial volume effects in the narrow sulcal regions, especially on the dHCP dataset. The surfaces predicted by Voxel2Mesh tend to be over-smoothed due to the mesh regularization terms.

\subsubsection{Self-intersection} To verify that CortexODE reduces surface self-intersections effectively, we compute the percentage of self-intersecting faces (SIFs) as shown in Table~\ref{table:comparison}. The results demonstrate that  CortexODE produces almost no SIFs in WM surfaces and less than 0.4\% SIFs in pial surfaces. CortexODE has significantly fewer SIFs than all other approaches, except that CorticalFlow has fewer SIFs in pial surfaces on the ADNI datasets. However, as illustrated in Figure \ref{fig:mri}, the sulcal regions are more narrow for neonates than adults. Therefore, CorticalFlow fails to prevent SIFs on the dHCP datasets, as it is difficult to deform a convex hull into deep and narrow sulci without intersections.
Although PialNN achieves similar geometric accuracy to CortexODE, it has much more SIFs than CortexODE in the pial surfaces as it has no theoretical guarantees to reduce SIFs, which impede its clinical applications (see Section \ref{sec:clinical-mesh}). DeepCSR introduces no SIFs since the surfaces are extracted by Marching Cubes.

For pial surface reconstruction, the input WM surfaces are inflated along the normal direction (Algorithm \ref{algo:inflate}). Therefore, our approach can also effectively reduce intersections between WM and pial surfaces. In our experiments, CortexODE only produces 1.57/1.24/1.77\% intersecting faces between WM and pial surfaces on the dHCP/HCP/ADNI datasets.

\begin{figure}[ht]
\centering
\includegraphics[width=1.0\linewidth]{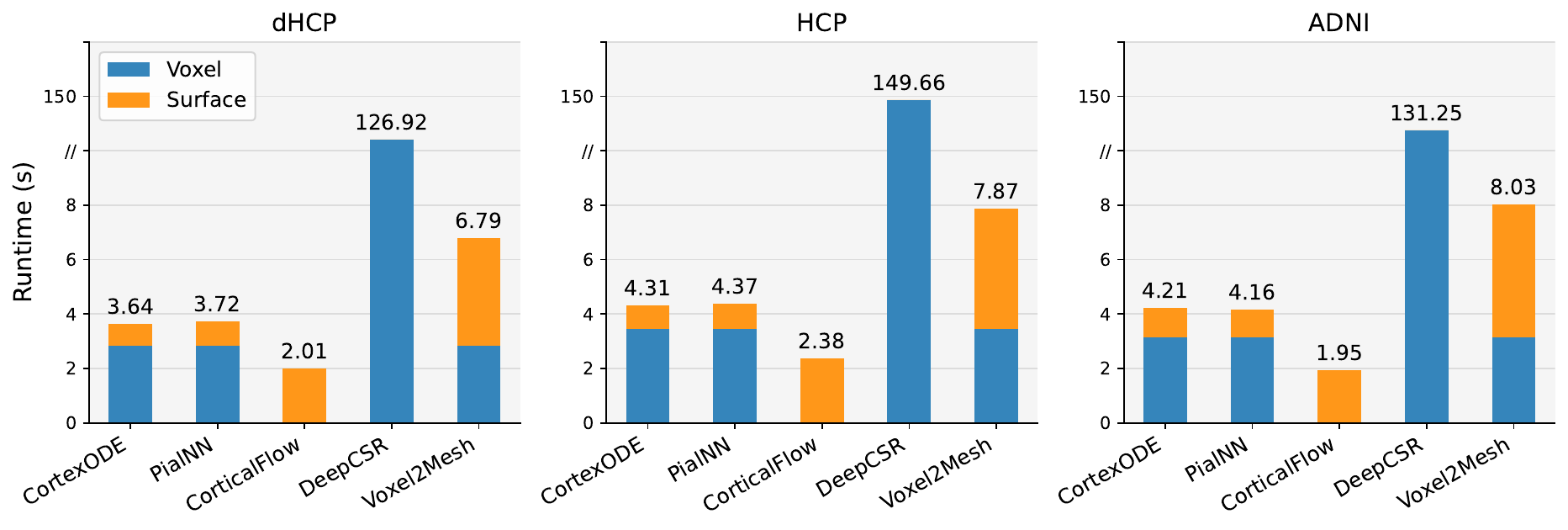}
\caption{The runtime ($second$) of cortical surface reconstruction for dHCP, HCP and ADNI datasets. The blue bars indicate the voxel-level processing time and the orange bars indicate surface-level runtime.}
\label{fig:runtime}
\end{figure}

\subsubsection{Runtime Analysis}
The inference time of all deep learning baselines is shown in Figure \ref{fig:runtime}. We divide the runtime into volume-level (\emph{e.g.} segmentation, topology correction and isosurface extraction) and surface-level (\emph{e.g.} mesh smoothing, subdividing and deformation) processing time. The runtime is measured on an Nvidia RTX3080 GPU and an Intel Core i7 CPU. From Figure \ref{fig:runtime}, since CortexODE, PialNN and Voxel2Mesh use the same pipeline to generate initial surfaces, they have the same volume-level processing time. The DeepCSR framework requires 2 minutes due to the voxel-wise prediction on a high resolution volume. CorticalFlow achieves the fastest runtime because it has no volume-level pre-processing. In summary, the CortexODE-based pipeline is accurate and efficient. It runs within 5 seconds for all datasets, whereas traditional pipelines~\cite{fischl2012freesurfer,glasser2013hcp,makropoulos2018dhcp} usually require more than 5 hours for a single subject's cortical surface reconstruction.

\subsection{ODE Integration}\label{sec:solver}
\subsubsection{ODE Solvers} CortexODE can be discretized by various explicit ODE solvers. We compare the performance of CortexODE models using different ODE solvers including forward Euler, midpoint (Mid), and fourth-order Runge Kutta (RK4). For evaluation, we train CortexODE models for pial surface reconstruction on the HCP dataset with different ODE solvers, including Euler-10 ($N=10$), Euler-20 ($N=20$), Mid-10 ($N=10$), and RK4-5 ($N=5$). $N$ is the number of total steps for ODE solvers in training and $h=T/N$ is the step size with $T=1s$. We set the inflation scale $\rho=0.001$ for the input WM surfaces to avoid the SIFs introduced by inflation. Several metrics of the ODE solvers for training are reported in Table~\ref{table:ode-solver}.
The Lipschitz constant $L_F$ of the deformation network $F_\theta$ is computed based on Eq.~(\ref{eq:L}) with 2-norm. The higher-order methods require more number of function evaluation (NFE), \emph{i.e.}, the times of forward propagation of $F_{\theta}$, but the error decreases much faster.
The training time increases linearly with the NFE, while the GPU memory cost is constant for training using adjoint method despite the step size.

\begin{table}[ht]
\centering
\caption{The metrics of ODE solvers for training. The training step size $h_{\mathrm{train}}$, number of function evaluation (NFE), accumulated error of ODE solvers, training time $t_{\mathrm{train}}$ for each backpropagation iteration, GPU memory cost for training, and Lipschitz constant $L_F$ are reported. }
\label{table:ode-solver}
\begin{tabular}{l|cccccc}
\toprule
Solver & $h_{\mathrm{train}}$ & NFE & Error & $t_{\mathrm{train}}$ & GPU & $L_F$\\
\midrule
Euler-10&  0.1   & 10 & $\mathcal{O}(h)$   & 0.47s/it & 7.15GB & 94.26\\
Euler-20&  0.05  & 20 & $\mathcal{O}(h)$   & 0.93s/it & 9.03GB & 95.72\\
Mid-10  &  0.1   & 20 & $\mathcal{O}(h^2)$ & 0.92s/it & 8.97GB & 110.10\\
RK4-5   &  0.2   & 20 & $\mathcal{O}(h^4)$ & 0.92s/it & 9.20GB & 107.14\\
\bottomrule
\end{tabular}
\end{table}

\subsubsection{Performance} 
\begin{figure}[ht]
\centering
\includegraphics[width=0.95\linewidth]{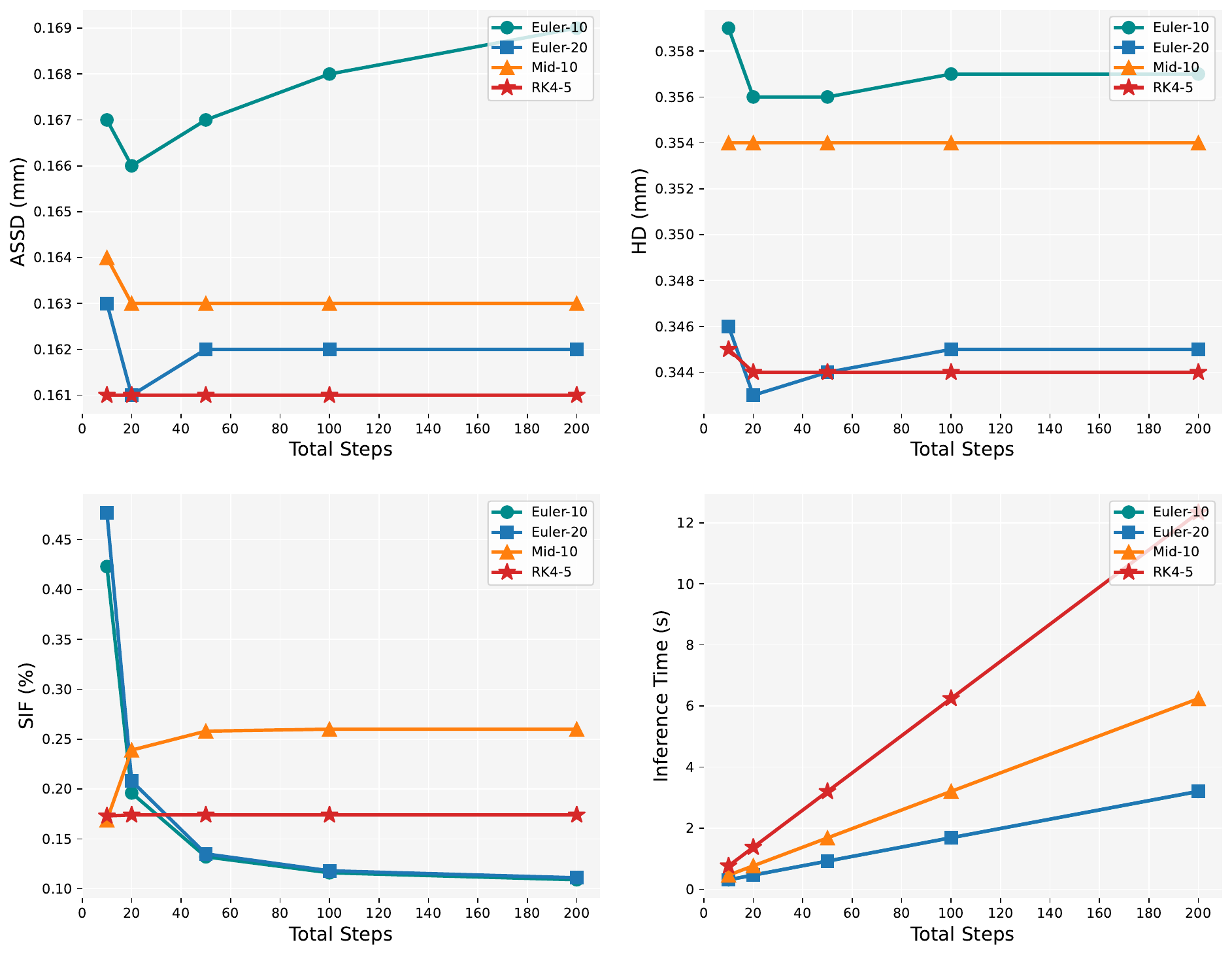}
\caption{Comparisons for different ODE solvers. We set fixed step sizes to train the CortexODE models and use different step sizes for the inference. The ASSD, HD, SIFs and inference time are evaluated for total steps $N=10,20,50,100,200$.}
\label{fig:ode-solver}
\end{figure}

The performance of the trained models is evaluated for total steps $N=10,20,50,100,200$ with step size $h=1/N$. The condition $\eta(h,L_F)<1$ in Theorem \ref{thm2} holds for all solvers when $N=200$. For each ODE solver, the ASSD, HD, percentage of SIFs and inference time are reported in Figure~\ref{fig:ode-solver}. For the lower-order Euler method, Figure~\ref{fig:ode-solver} reveals that  by decreasing the step size, the ratio of SIFs reduces to 0.11\% but the geometric error increases slightly.
The higher-order methods such as midpoint and RK4 provide more accurate and stable approximation for ODE solutions, whereas they are more computationally intensive. The performance will not decrease when tested on small step sizes, while the SIFs do not decrease as well. In practice, we can choose suitable ODE solvers and step sizes for different scenarios determined by the demand for accuracy, efficiency and mesh quality.

\subsubsection{Limitations} One limitation of CortexODE is that the SIFs still exist even though the condition $\eta(h,L)<1$ is already satisfied. As discussed in Section \ref{sec:ode-integrate}, one reason is that the homeomorphic mapping should be applied to all points on a smooth surface to prevent self-intersections. However, the polygon mesh only has a limited number of vertices. This inevitable discretization causes resolution limits and thus self-intersections.
In addition, the CortexODE model is trained using a  relatively large step size $h$, which produces SIFs during training. This also introduces SIFs in the inference despite  using of smaller step sizes, especially for higher-order solvers which are more stable. It should be noted that although SIFs are not completely prevented, CortexODE still reduces SIFs effectively and achieves better mesh quality than existing learning-based approaches while providing theoretical support.

\subsection{Clinical Analysis}
\subsubsection{Reliability}
The reliability of cortical surface reconstruction is an important feature required by clinical applications. Based on previous work~\cite{henschel2020fastsurfer,cruz2020deepcsr}, we evaluate the reliability of CortexODE on a Test-Retest dataset~\cite{maclaren2014reliability}. The Test-Retest dataset contains 3 subjects, each of which has 40 brain MRI scans collected within a short period. Hence, the cortical surfaces of the same subject should be identical or highly consistent. 
We use models pre-trained on the ADNI dataset to extract the cortical surfaces from all brain MRI volumes. The ASSD and HD are measured between all pairs of the surfaces for each subject, where each pair of the surfaces are aligned by the iterative closest point (ICP) algorithm. The results including statistical significance (t-test) are reported in Table~\ref{table:test-retest}. It shows that CortexODE achieves the best consistency between cortical surfaces compared to FreeSurfer and existing learning-based approaches.

\begin{table}[ht]
\centering\setlength\tabcolsep{4.5pt}
\caption{Comparative results of test-retest reliability. The ASSD and HD are measured between each pair of cortical surfaces on the Test-Retest dataset. *Significant difference ($p<0.05$) compared to CortexODE.} \label{table:test-retest}
\begin{tabular}{l|ccc|ccc}
\toprule
\multicolumn{1}{l}{} & \multicolumn{3}{c}{ASSD ($mm$)} & \multicolumn{3}{c}{HD ($mm$)} \\
\midrule
Method & M & SD & $p$ &  M & SD & $p$ \\
\midrule
CortexODE & \textbf{0.208} & 0.031 & --- & \textbf{0.440} & 0.073 & ---\\
FreeSurfer & 0.225 & 0.024 &  $<$0.001* & 0.476 & 0.058 &  $<$0.001*\\
PialNN & 0.213 & 0.031 & $<$0.001* & 0.445 & 0.070 & 0.038*\\
CorticalFlow & 0.220 & 0.033 & $<$0.001* & 0.464 & 0.072 &  $<$0.001*\\
DeepCSR & 0.268 & 0.063 & $<$0.001* & 0.573 & 0.148 & $<$0.001*\\
Voxel2Mesh & 0.323 & 0.066 & $<$0.001* & 0.694 & 0.153 & $<$0.001*\\
\bottomrule
\end{tabular}
\end{table}

\subsubsection{Surface Quality}\label{sec:clinical-mesh}
The quality of the cortical surfaces can affect the estimation of cortical features which are crucial for downstream clinical analysis~\cite{fischl2000measuring,rimol2012clinical}. We evaluate the impact of mesh self-intersections on the measurement of cortical thickness, (pial) surface area and cortical volume. We train a CortexODE model on the HCP dataset using an Euler-20 solver. The geometric accuracy and ratio of SIFs of pial surfaces, as well as the cortical features are reported in Table \ref{table:mesh-quality} as the \emph{Baseline} to compare. We set the time step $h$=0.01 for inference so that the results only have 0.118\% SIFs. Next, we introduce two types of SIFs into the cortical surfaces without changing geometric accuracy. This provides fair comparisons to explore the effect of SIFs on the cortical features.

As a first type, we consider the general SIFs that are caused by poor mesh quality. Given pial surfaces predicted by CortexODE, we randomly flip the edge of the mesh, \emph{i.e.}, flip the coordinates between any pair of adjacent vertices to produce extra SIFs in the surfaces. We flip 0.4\% of the edges for each surface and evaluate the geometric errors (ASSD and HD) as well as the cortical features. We use a t-test to examine the significance between \emph{Baseline} and \emph{Flipping}. Table \ref{table:mesh-quality} shows that the flipping creates 3.25\% SIFs but does not show significant changes in distance-based geometric errors. However, the surface area is significantly over-estimated due to the intersecting and overlapping between the mesh faces.

As a second type, for cortical surfaces, SIFs often occur in the deep and narrow sulci for pial surfaces due to partial volume effects~\cite{ballester2002partial}. To investigate this, we use a large integration time step $h=0.1$ for CortexODE (\emph{Large Step}) such that $\eta(h,L)>1$ and SIFs in the sulci cannot be prevented. The results in Table~\ref{table:mesh-quality} indicate that, although no significant difference is found in geometric accuracy, the cortical thickness and volume of \emph{Large Step} are increased significantly compared to the \emph{Baseline}. This results from the increase of SIFs (0.477\%) which occur along with the over-inflation of the cortical surfaces in sulcal regions.

According to above experiments, even though for cortical surfaces with the same geometric accuracy, the occurring of SIFs can significantly affect the estimation of cortical features. Therefore, it is essential to prevent surface self-intersections to provide reliable cortical features for clinical analysis.

\begin{table}[ht]
\setlength\tabcolsep{2.8pt}
\centering
\caption{The results of cortical feature estimation. The ASSD ($mm$), HD ($mm$), ratio of SIFs ($\%$), cortical thickness ($mm$), surface area ($cm^2$) and cortical volume ($cm^3$) for one brain hemisphere are reported. The results are bold if there are significant differences ($p<0.05$*) compared to Baseline.}\label{table:mesh-quality}
\begin{tabular}{l|cc|ccc|ccc}
\toprule
\multicolumn{1}{l}{} & \multicolumn{2}{c}{Baseline}& \multicolumn{3}{c}{Flipping} & \multicolumn{3}{c}{Large Step} \\
\midrule
Metrics & M & SD & M & SD & $p$ & M & SD & $p$ \\
\midrule
ASSD & 0.162  & 0.016 & 0.163 & 0.016  & 0.452 & 0.163 & 0.018 & 0.728    \\
HD & 0.345  & 0.035 & 0.346 & 0.035  & 0.699 & 0.346 & 0.039 & 0.678    \\
SIF & 0.118  & 0.064 & \textbf{3.249} & \textbf{0.071}  & $<$0.001* & \textbf{0.477} & \textbf{0.255} & $<$0.001*    \\
Thickness & 2.566  & 0.068 & 2.566 & 0.068  & 0.998 & \textbf{2.597} & \textbf{0.068} & $<$0.001*   \\
Area   & 1079  & 113 & \textbf{1106} & \textbf{116}  & 0.002*  & 1087 & 113 & 0.379    \\
Volume  & 257  & 29 & 257 & 29 & 0.910 & \textbf{262} & \textbf{29}  & 0.030*   \\
\bottomrule
\end{tabular}
\end{table}

\subsection{Ablation Study}
We conduct ablation experiments to verify the effectiveness of the components in the CortexODE-based pipeline.
\subsubsection{Segmentation}\label{sec:ablation-seg}
We fist investigate how different segmentation approaches can affect the geometric accuracy of cortical surfaces extracted by CortexODE. We compare the performance of WM segmentation between U-Net (U-Net-Seg)~\cite{ronneberger2015unet}, QuickNAT~\cite{roy2019quicknat} and FastSurfer \cite{henschel2020fastsurfer} on the HCP dataset. The Dice coefficient, Intersection over Union (IoU), and runtime of the approaches are reported in Table \ref{table:segmentation}, where the runtime consists of the inference time of neural networks and post-processing time of implicit surface creation. The geometric errors of the reconstructed cortical surfaces using the predicted segmentations are reported in Table \ref{table:segmentation} as well. 

In addition, we consider using a U-Net (U-Net-SDF) to predict the SDF directly. The WM segmentation of U-Net-SDF is defined by the voxels within 0-level. Table \ref{table:segmentation} shows that QuickNAT and FastSurfer can provide better segmentation, which results in more accurate cortical surfaces for CortexODE. However, both of them are computationally intensive and require 4-8 seconds for inference. U-Net-SDF performs worse than U-Net-Seg since it is more difficult to learn a distance-based level set than a binary segmentation.

\begin{figure}[ht]
\centering
\includegraphics[width=0.95\linewidth]{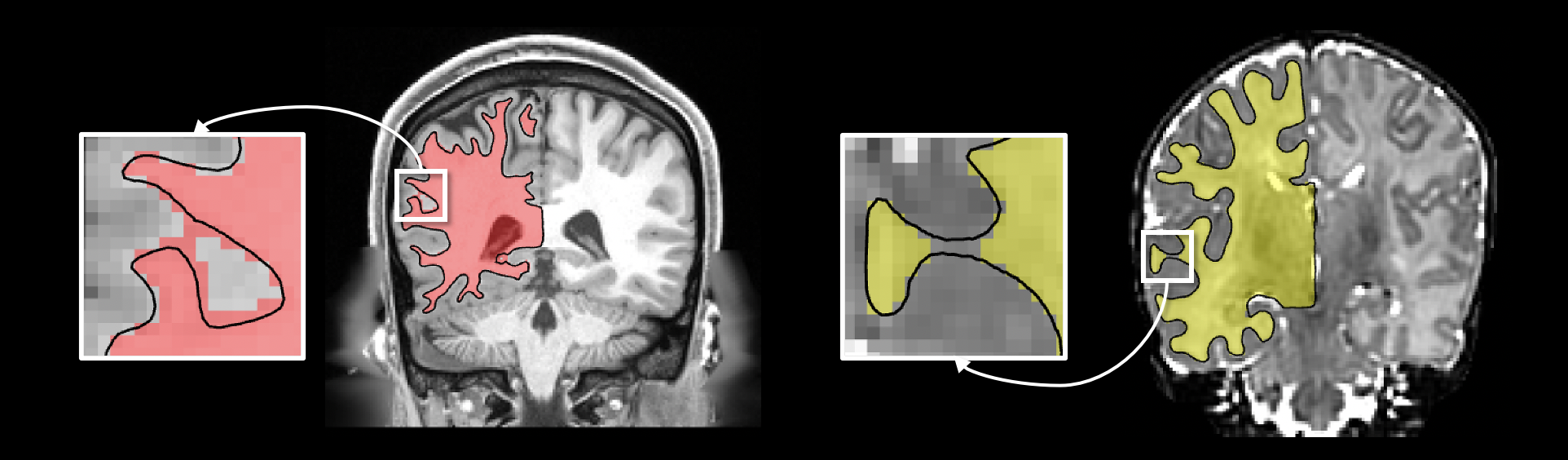}
\caption{Cortical surfaces (black contour) extracted by CortexODE pipeline based on imperfect segmentations. Left: Over-segmentation on the HCP data. Right: Disconnected segmentation on the dHCP data.}
\label{fig:seg}
\end{figure}

Learning-based approaches can produce imperfect initial segmentations. Figure \ref{fig:seg} shows the cortical surfaces predicted by CortexODE when the WM segmentation is over-segmented or disconnected. It demonstrates that our CortexODE pipeline is robust and can reconstruct accurate cortical surfaces even for cases in which the WM segmentation is inaccurate.

\begin{table}[ht]
\setlength\tabcolsep{4.0pt}
\centering
\caption{Ablation study for WM segmentation in CortexODE pipeline. The Dice coefficient ($\%$), IoU ($\%$) and runtime ($second$) for segmentation are reported. The ASSD and HD are compared between the cortical surfaces extracted by CortexODE pipeline using predicted segmentations.}
\label{table:segmentation}
\begin{tabular}{l|ccc|cc|cc}
\toprule
\multicolumn{1}{l}{} & \multicolumn{3}{c}{Segmentation}& \multicolumn{2}{c}{WM Surface} & \multicolumn{2}{c}{Pial Surface}\\
\midrule
Method & Dice & IoU & Runtime & ASSD & HD & ASSD & HD\\
\midrule
U-Net-Seg  & 98.07 & 96.21 & 0.596 & 0.115 & 0.233 & 0.163 & 0.347\\
U-Net-SDF  & 97.12 & 94.43 & \textbf{0.235} & 0.146 & 0.296 & 0.165 & 0.349\\
QuickNAT   & \textbf{98.57} & \textbf{97.18} & 8.342 & \textbf{0.103} & \textbf{0.208} & \textbf{0.155} & 0.332 \\
FastSurfer & 98.52 & 97.09 & 4.643 & 0.104 & \textbf{0.208} & \textbf{0.155} & \textbf{0.331} \\
\bottomrule
\end{tabular}
\end{table}

\subsubsection{Topology Correction}\label{sec:topology}
We compare our accelerated topology correction algorithm to its original implementation~\cite{bazin2007topology}. The runtime is measured on an Intel Core i7 CPU. As shown in Table~\ref{table:topology}, for discrete SDFs with size of $192\times224\times192$ on the HCP dataset, the Numba-accelerated algorithm with improved initialization (Numba+Init) only runs one second on average. This is $20\times$ faster than the original Java implementation which requires 21.48s. The Numba compiler accelerates the Python implementation from 144.38s to 3.45s. Our initialization method reduces the number of processed voxels by 76$\%$ for topology correction, and thus it speeds up the original algorithm effectively. In all implementations, the corrected SDFs produce surfaces with the Euler characteristic $\chi=V-E+F=2$, which means the surfaces have genus zero topology. We also check the topology of cortical surfaces generated by the baseline models in Table \ref{table:comparison}. The average Euler characteristic number is 2 for all approaches, which means all predicted surfaces have the correct topology.

\begin{table}[ht]
\centering
\caption{Comparisons for different implementations of the topology correction algorithm. The runtime, percentage of processed voxels and Euler characteristic number are reported. The results are evaluated on the discrete SDFs of left WM surfaces with size of $192\times224\times192$.}
\label{table:topology}
\begin{tabular}{l|ccc}
\toprule
Method & Runtime & Processed Voxels & Euler Number\\
\midrule
Numba+Init  &  0.99s   & 20.71$\%$ & 2 \\
Numba       &  3.45s   & 97.05$\%$ & 2 \\
Python+Init &  27.78s  & 20.71$\%$ & 2  \\
Python      &  144.38s & 97.05$\%$ & 2  \\
Java (Original)  &  21.48s  & 97.05$\%$ & 2  \\
\bottomrule
\end{tabular}
\end{table}

\subsubsection{Initial Surface Extraction}

\begin{figure}[ht]
\centering
\includegraphics[width=0.95\linewidth]{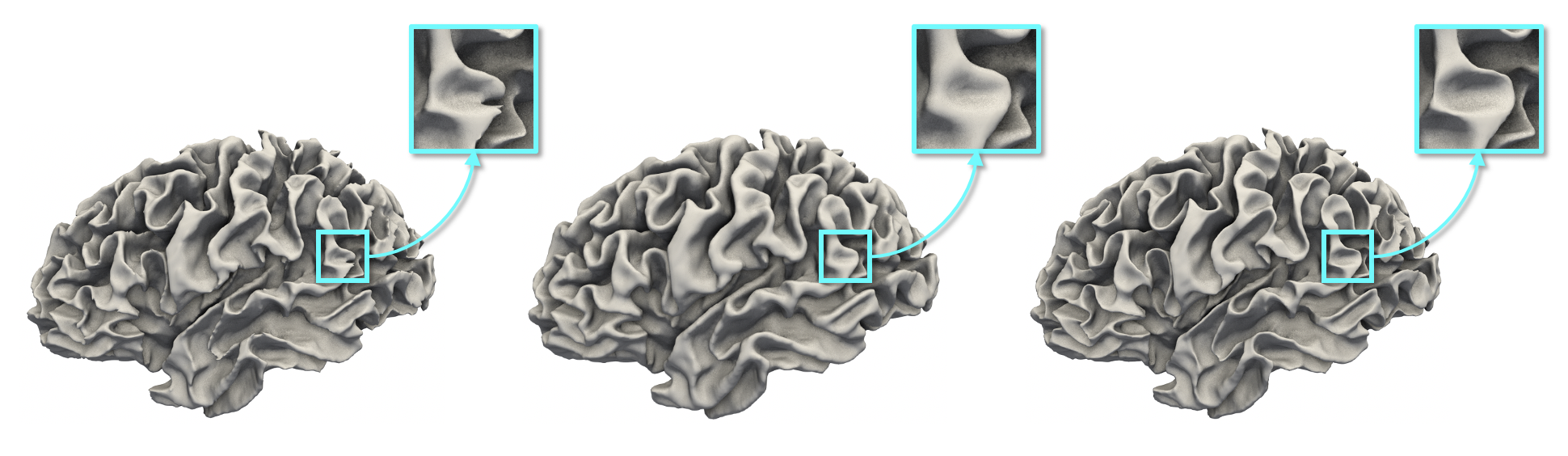}
\caption{Left: Implicit WM surface defined by SDF (0-level surface). Middle: Initial surface extracted at $\lambda_0$-level. Right: WM surface predicted by CortexODE.}
\label{fig:init-surf}
\end{figure}

\begin{figure}[ht]
\centering 
\includegraphics[width=0.95\linewidth]{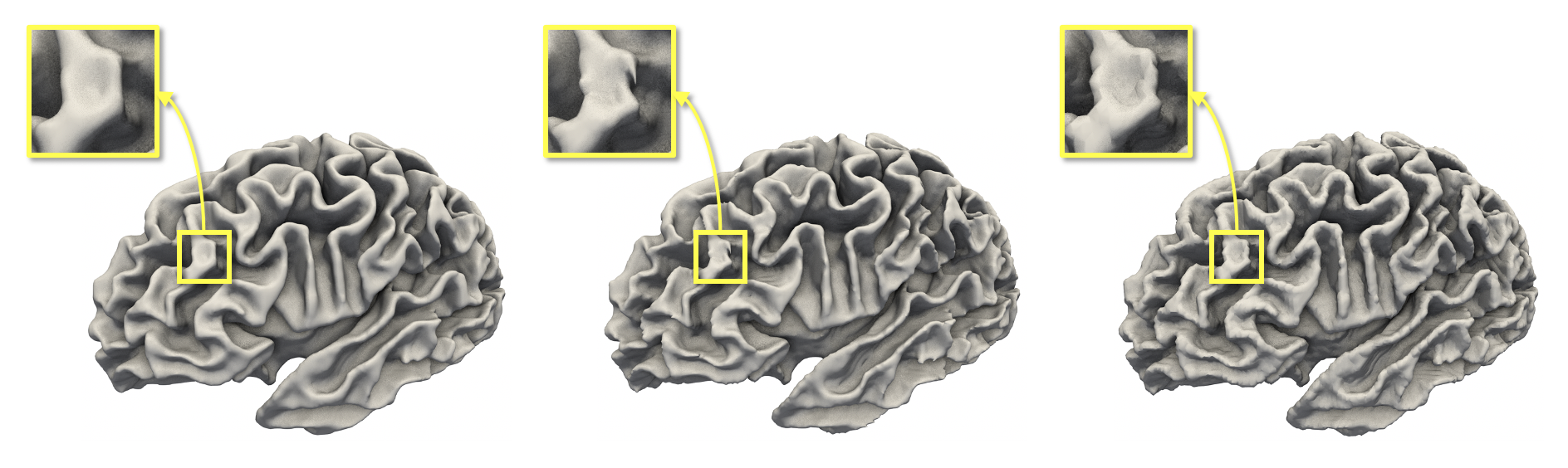}
\caption{ Left: Initial surface created by CortexODE. Middle: Initial surface without implicit surface smoothing. Right: Initial surface without Laplacian mesh smoothing.}
\label{fig:smooth}
\end{figure}

We verify the effectiveness of the processing pipeline for initial surface extraction. More specifically, we conduct ablation studies on isosurface extraction and surface smoothing. 

For isosurface extraction, we employ the Marching Cubes algorithm~\cite{lorensen1987marching} to extract an inflated WM surface as the initial surface at $\lambda_0$-level of the SDF instead of 0-level. As shown in Figure \ref{fig:init-surf}-Left, the implicit WM surface defined by the SDF (0-level surface) is corrupted due to the imperfect segmentation, while the initial surface at $\lambda_0$-level (Figure \ref{fig:init-surf}-Middle) can reduce these corruptions effectively. Generally $-1<\lambda_0<-0.5$ is a reasonable range for the initial surface extraction. In our experiments, we set $\lambda_0=-0.8$ which is validated to achieve superior results in various age groups including the dHCP, HCP and ADNI datasets (Table \ref{table:comparison}). For the HCP dataset, the geometric error (ASSD) is 0.607$mm$ between the initial surfaces and GT WM surfaces. The CortexODE model deforms the initial surfaces into WM surfaces (Figure \ref{fig:init-surf}-Right) and reduces the error to ASSD=0.115$mm$. This performs much better than the implicit WM surface at 0-level which only achieves ASSD=0.211$mm$.

The CortexODE pipeline involves two smoothing steps. First, an implicit surface smoothing with a Gaussian filer is applied to the SDF to reduce the corruptions caused by inaccurate segmentation. Second, two iterations of Laplacian mesh smoothing are used to remove the artifacts introduced by Marching Cubes. Figure \ref{fig:smooth} visualizes the initial surfaces without either of two smoothing steps. It turns out that the smoothing steps are necessary to produce high-quality surface meshes. In practice, the Gaussian blurring with standard deviation $\sigma=0.5$ is sufficient to avoid most of the corruptions and keep the details of the surface.

\begin{figure}[ht]
\centering
\includegraphics[width=.95\linewidth]{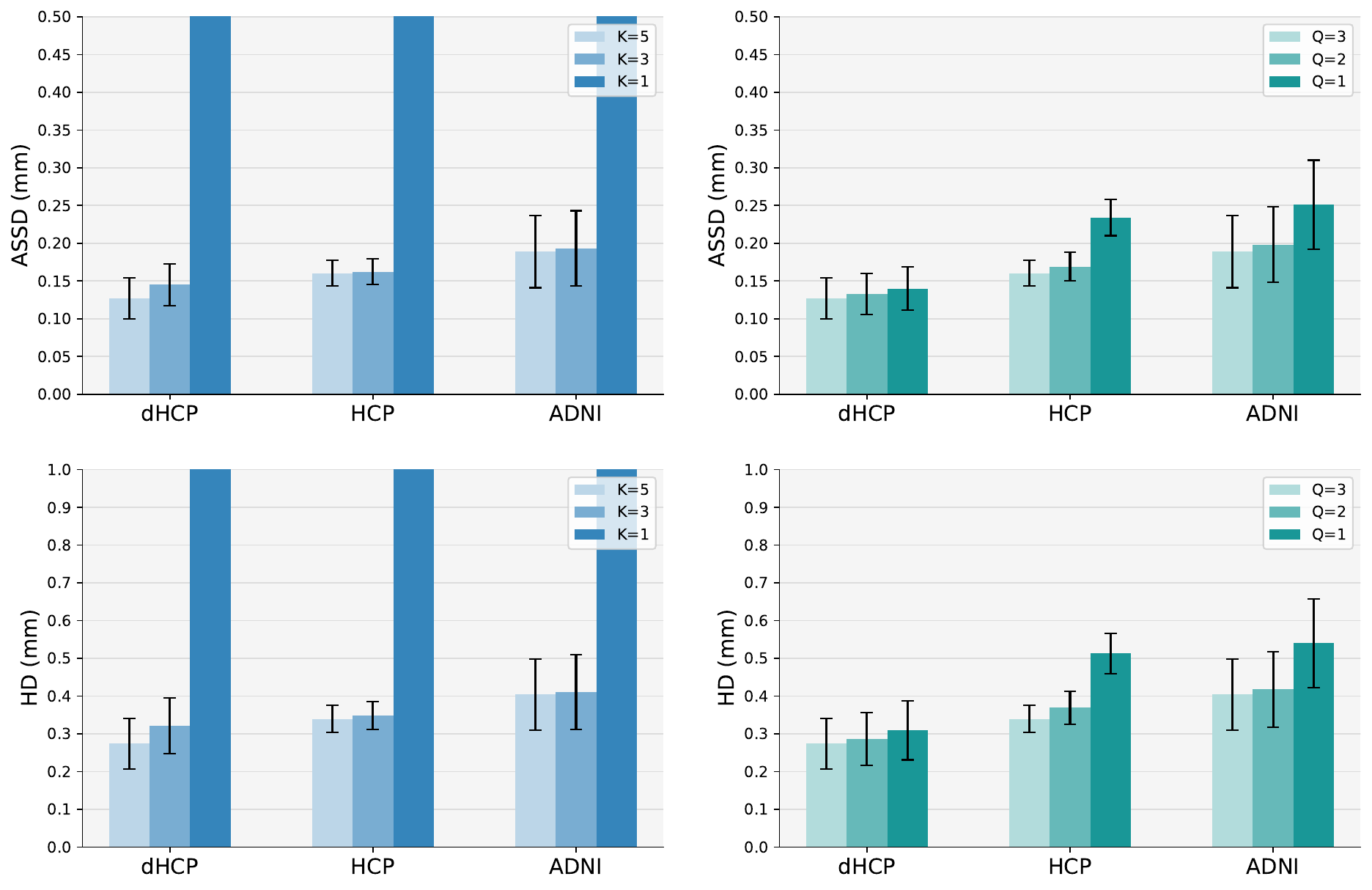}
\caption{Ablation study for CortexODE architecture. The ASSD and HD of the reconstructed pial surfaces are reported for input scales $Q=1,2,3$ and cube sizes $K=1,3,5$.}
\label{fig:ablation-arch}
\end{figure}

\subsubsection{CortexODE Architecture} To examine the architecture of the CortexODE framework, we consider the effect of different input scales and cube sampling sizes. We train CortexODE models with input scales $Q=1,2,3$ or cube sizes $K=1,3,5$ for pial surface reconstruction on dHCP, HCP and ADNI datasets. The ASSD and HD are provided in Figure~\ref{fig:ablation-arch}, which demonstrates that multiple scales and large cube size improve the performance of CortexODE effectively. A larger kernel/cube size $K$ will enlarge the receptive fields, which lead to better performance but also increase the GPU memory requirements. The selection of scale $Q$ depends on the resolution of the input brain MRI image. We can use $Q=2$ for low-resolution image and $Q=3$ for higher resolution. Generally, it is sufficient to use $K=5$ and $Q=3$ to produce accurate cortical surfaces.

\section{Conclusion}\label{sec:conclusion}
In this work, we introduce CortexODE, a novel geometric deep learning framework for cortical surface deformation. CortexODE defines a diffeomorphic flow between surfaces and provides theoretical guarantees to prevent surface self-intersections. An efficient CortexODE-based pipeline is further developed by incorporating WM segmentation, distance transform and fast topology correction. The experiments demonstrate that the proposed pipeline reconstructs high quality cortical surfaces in less than 5 seconds with $<$0.2$mm$ geometric error and $<$0.4\% self-intersecting faces. Our pipeline provides a fast alternative solution for large-scale neuroimage studies in various age groups including neonates and adults.

\appendices
\section{Proof of Theorems}
\subsection{Theorem 1: Existence and Uniqueness of Solutions}\label{proof:thm1}
\begin{proof}
For FC layer $f_l$ and any $\mathbf{x},\mathbf{y}$, it holds
\begin{equation}\label{eq:lip-fc}
\|f_l(\mathbf{x})-f_l(\mathbf{y})\|_p
\leq L_{\phi}\|W_l\|_p  \|\mathbf{x}-\mathbf{y}\|_p,
\end{equation}
where $L_{\phi}=1$ for ReLU or LeakyReLU activation functions. Based on Eq. (\ref{eq:F}) and (\ref{eq:lip-fc}), we have
\begin{equation}\label{eq:lip-F}
\begin{split}
&\|F_{\theta}(\mathbf{x},\mathbf{V})-F_{\theta}(\mathbf{y},\mathbf{V})\|_p \leq \|W_5\|_p\|W_4\|_p\|W_3\|_p\cdot\\
&\left(\|W_1\|_p^p\|\mathbf{x}-\mathbf{y}\|_p^p
+\|W_2\|_p^p \|W_0\|_p^p \|\mathbf{v}(\mathbf{x})-\mathbf{v}(\mathbf{y})\|_p^p\right)^{\frac{1}{p}}
\end{split}
\end{equation}
By the definition of $\mathbf{v}(\mathbf{x})$ in Eq.~(\ref{eq:v_qijk}),
\begin{equation}\label{eq:lip-v}
\begin{split}
&\|\mathbf{v}(\mathbf{x})-\mathbf{v}(\mathbf{y})\|_p^p=\sum_{q=1}^Q \sum_{i,j,k=1}^K|\mathbf{v}_{qijk}(\mathbf{x})-\mathbf{v}_{qijk}(\mathbf{y})|^p\\
&\leq K^3 \sum\nolimits_{q=1}^Q 2^{p(1-q)}((\mathbf{V}_q)_{\mathrm{max}}-(\mathbf{V}_q)_{\mathrm{min}})^p\|\mathbf{x}-\mathbf{y}\|_p^p\\
&\leq 2K^3 (\mathbf{V}_{\mathrm{max}}-\mathbf{V}_{\mathrm{min}})^p\|\mathbf{x}-\mathbf{y}\|_p^p.\\
\end{split}
\end{equation}
The first inequality holds because the trilinear interpolation $\mathrm{Lerp}(\cdot)$ is piece-wise linear and continuous.

Combining Eq.~(\ref{eq:lip-F}) and (\ref{eq:lip-v}), we can conclude that $F_{\theta}$ is Lipschitz continuous, \emph{i.e.}, 
\begin{equation}
\|F_{\theta}(\mathbf{x},\mathbf{V})-F_{\theta}(\mathbf{y},\mathbf{V})\|_p \leq L_F \|\mathbf{x}-\mathbf{y}\|_p,
\end{equation}
where $L_F$ is defined in Eq.~(\ref{eq:L}). Since $F_{\theta}$ is the RHS function of ODE~(\ref{eq:cortexode}), according to the existence and uniqueness theorem of ODE solutions~\cite{coddington1955ode}, there exists a unique solution for the IVP in Eq.~(\ref{eq:cortexode}).
\end{proof}

\subsection{Theorem 2: Homeomorphism of ODE Integration}\label{proof:thm2}
\begin{proof}
Firstly, by the Lipschitz continuity of $F_{\theta}$, we have
\begin{equation}\label{eq:si}
\begin{split}
\|s_i(\mathbf{x})-s_i(\mathbf{y})\|_p & \leq  L\|\mathbf{x}-\mathbf{y}\|_p \\ &+hL\sum\nolimits_{j=1}^{i-1}|\alpha_{i,j}|\|s_j(\mathbf{x})-s_j(\mathbf{y})\|_p.
\end{split}
\end{equation}
Following the proofs in \cite{lebrat2021corticalflow}, for any $\mathbf{x},\mathbf{y}$ it holds
\begin{equation}\label{eq:injection}
\begin{split}
&\|\mathcal{G}(\mathbf{x})-\mathcal{G}(\mathbf{y})\|_p
\geq\|\mathbf{x}-\mathbf{y}\|_p-h\sum_{i_0=1}^r|\beta_{i_0}|\|s_{i_0}(\mathbf{x})-s_{i_0}(\mathbf{y})\|_p\\
&\geq (1-hL\sum\nolimits_{i_0=1}^r|\beta_{i_0}|)\|\mathbf{x}-\mathbf{y}\|_p\\
&~~~~~~~~~~~~ -h^2L\sum_{i_0=1}^r|\beta_{i_0}|\sum_{i_1=1}^{i_0-1}|\alpha_{i_0,i_1}|\|s_{i_1}(\mathbf{x})-s_{i_1}(\mathbf{y})\|_p\\
&\geq (1-hL\sum_{i_0=1}^r|\beta_{i_0}|-h^2L^2\sum_{i_0=1}^r|\beta_{i_0}|\sum_{i_1=1}^{i_0-1}|\alpha_{i_0,i_1}|)\|\mathbf{x}-\mathbf{y}\|_p\\
&-h^3L^2\sum_{i_0=1}^r|\beta_{i_0}|\sum_{i_1=1}^{i_0-1}|\alpha_{i_0,i_1}|\sum_{i_2=1}^{i_1-1} |\alpha_{i_1,i_2}| \|s_{i_2}(\mathbf{x})-s_{i_2}(\mathbf{y})\|_p\\
&\geq \left(1-\eta(h,L)\right)\|\mathbf{x}-\mathbf{y}\|_p.\\
\end{split}
\end{equation}
Since $\eta(h,L)<1$, we have $\mathcal{G}(\mathbf{x})\neq\mathcal{G}(\mathbf{y})$ for any $\mathbf{x}\neq\mathbf{y}$, which means $\mathcal{G}$ is an injection.

To prove $\mathcal{G}$ is a surjection, for any $\mathbf{z}$, we define $\mathcal{H}(\mathbf{x})=\mathbf{z}-h\sum_{i=1}^r\beta_i s_i(\mathbf{x})$. Similar to the procedure in (\ref{eq:injection}), for any $\mathbf{x}$ and $\mathbf{y}$,
\begin{equation}\label{eq:surjection}
\begin{split}
\|\mathcal{H}(\mathbf{x})-\mathcal{H}(\mathbf{y})\|_p&=\|h\sum\nolimits_{i=1}^r\beta_{i}(s_{i}(\mathbf{x})-s_{i}(\mathbf{y}))\|_p\\
&\leq \eta(h,L)\|\mathbf{x}-\mathbf{y}\|_p.
\end{split}
\end{equation}
According to the contraction mapping theorem, there exists a unique fixed point $\bar{\mathbf{x}}$ such that $\mathcal{H}(\bar{\mathbf{x}})=\bar{\mathbf{x}}$. Hence, for any $\mathbf{z}$, we can find a unique $\bar{\mathbf{x}}$ such that $\mathcal{G}(\bar{\mathbf{x}}) = \mathbf{z}$, \emph{i.e.}, $\mathcal{G}$ is a surjection.

Similar to (\ref{eq:injection}) and (\ref{eq:surjection}), $\mathcal{G}$ is Lipschitz continuous with Lipschitz constant $1+\eta(h,L)$. By replacing $\mathbf{x}$ with $\mathcal{G}^{-1}(\mathbf{x})$ in (\ref{eq:injection}), we show that $\mathcal{G}^{-1}$ is also Lipschitz continuous:
\begin{equation}
\begin{split}
&\|\mathcal{G}^{-1}(\mathbf{x})-\mathcal{G}^{-1}(\mathbf{y})\|_p\leq(1-\eta(h,L))^{-1}\|\mathbf{x}-\mathbf{y}\|_p.
\end{split}
\end{equation}
Therefore, $\mathcal{G}$ is a homeomorphism.
\end{proof}






\bibliographystyle{ieeetr}
\bibliography{ref}
\end{document}